\documentclass[lettersize,journal]{IEEEtran}
\IEEEoverridecommandlockouts

\usepackage{cite}
\usepackage{booktabs}
\usepackage{lipsum}
\usepackage{xurl}
\usepackage{multicol}
\usepackage{amsmath,amssymb,amsfonts}
\usepackage{enumitem}
\usepackage[official]{eurosym}
\usepackage{algorithm,algpseudocode}
\usepackage{xcolor}
\usepackage{graphicx}
\usepackage{setspace}
\usepackage{float}
\usepackage{caption}
\usepackage{subcaption}
\usepackage{supertabular}
\usepackage{multirow}
\usepackage{footnote}
\makesavenoteenv{tabular}
\makesavenoteenv{table}
\usepackage[perpage,hang,flushmargin]{footmisc}
\usepackage{url}
\def\BibTeX{{\rm B\kern-.05em{\sc i\kern-.025em b}\kern-.08em
    T\kern-.1667em\lower.7ex\hbox{E}\kern-.125emX}}
\renewcommand{\footnotesize}{\scriptsize}

\newcommand{\etal}{\textit{et al.}}

\begin{document}

\title{When Computing follows Vehicles: Decentralized Mobility-Aware Resource Allocation for Edge-to-Cloud Continuum}
\author{Zeinab Nezami,
\thanks{Zeinab Nezami is a Postdoctoral Researcher at the School of Computing,
University of Leeds, Leeds, United Kingdom (e-mail: z.nezami@leeds.ac.uk).}
Emmanouil Chaniotakis, and
\thanks{Emmanouil Chaniotakis is an Associate Professor with the UCL Energy Institute, University College London, London, United Kingdom (e-mail: m.chaniotakis@ucl.ac.uk).}
Evangelos Pournaras
\thanks{Evangelos Pournaras is a Professor at the School of Computing,
University of Leeds, Leeds, United Kingdom (e-mail: e.pournaras@leeds.ac.uk).}}

\maketitle

\begin{abstract}
The transformation of smart mobility is unprecedented--Autonomous, shared and electric connected vehicles, along with the urgent need to meet ambitious net-zero targets by shifting to low-carbon transport modalities result in new traffic patterns and requirements for real-time computation at large-scale, for instance, augmented reality applications. The cloud computing paradigm can neither respond to such low-latency requirements nor adapt resource allocation to such dynamic spatio-temporal service requests. This paper addresses this grand challenge by introducing a novel decentralized optimization framework for mobility-aware edge-to-cloud resource allocation, service offloading,  provisioning and load-balancing. In contrast to related work, this framework comes with superior efficiency and cost-effectiveness under evaluation in real-world traffic settings and mobility datasets. This breakthrough capability of `\emph{computing follows vehicles}' proves able to reduce utilization variance by more than 40 times, while preventing service deadline violations by 14\%-34\%. 

\end{abstract}

\begin{IEEEkeywords}
Edge-to-Cloud Computing, Smart Mobility, Distributed Optimization, Dynamic Resource Allocation, Multi-agent System. 
\end{IEEEkeywords}

\section{Introduction}

\IEEEPARstart{I}{n} 2019, the UK domestic transport emerged as the largest emitting sector of greenhouse gases, accounting for 27\% of the UK's total emissions\footnote{Transport energy and environment statistics, available at \url{https://www.gov.uk/government/statistics/transport-and-environment-statistics-2021} (accessed April 2024).}, underscoring the need to prioritize CO2 emission reduction via traffic management and implementation of smart mobility services. With the advancements in Intelligent Transportation System (ITS), technologies such as connected and autonomous vehicles are envisioned to provide a safer, environmentally friendly, and convenient~\cite{chen2017service,ligo2017throughput} transportation ecosystem for the public. Vehicles are equipped with wireless communication capabilities to support a plethora of applications including rerouting, road safety, and location-based services~\cite{lu2014connected,lin2017resource}. 
The imminent proliferation of Internet of Things (IoT) devices, including vehicles, are projected to surpass 55.7 billion by 2025 and generate up to 79.4 zettabytes of data~\cite{rydning2018digitization}. The upsurge in data generation coincides with the considerable energy consumption attributable to ICT, already surpassing 10\% of global energy consumption and forecasted to exceed 20\% by 2030~\cite{jones2018stop}. These developments underscore the challenge of Information and Communication Technology (ICT) in responding and adapting, a novel concern overlooked within the computing-mobility ecosystem and addressed in this paper. 
\par The surge in data generation is compounded by the mobility of vehicles, a pivotal characteristic within the Internet of Things (IoT) environment, introducing dynamism and uncertainty. This dynamic nature poses numerous challenges, particularly in ICT resource management for applications with stringent Quality of Service (QoS) requirements. Current Vehicle-to-Cloud (V2C) and Vehicle-to-Vehicle (V2V) architectures struggle to meet diverse mobility service requirements, especially in terms of latency, efficiency, and scalability~\cite{chen2017empirical}. Vehicle-to-Infrastructure (V2I) enhances data distribution effectiveness by shifting cloud services to network edges. Fog and edge computing~\cite{bonomi2014fog,li2019energy,Nezami2019Internet} present a novel paradigm for offloading computation from vehicles to local servers while facilitating collaborative data sharing~\cite{satria2017recovery} and traffic rerouting~\cite{Gerostathopoulos2019}. The distributed architecture of fog computing (i.e., edge-to-cloud) reduces data traversal through the network, resulting in significant energy savings of 14\% to over 80\% compared to fully centralized cloud center architectures~\cite{ahvar2019estimating,yan2019modeling}.

\par This paper introduces a novel service provisioning framework tailored to address the evolving mobility ecosystem. The distributed, heterogeneous, and resource-constrained nature of fog infrastructure presents challenges in hosting and managing (i.e., service placement) the dynamic and stochastic nature of vehicular networks. Concentrating smart mobility services at nodes near intersections or vehicle destinations to meet demanding QoS requirements may lead to server overloading, inefficient resource distribution, and potential over-provisioning for some services, while others remain under-provisioned. In addition to workload balance and QoS parameters, various optimization aspects must be considered to effectively tackle the service provisioning problem, especially given the presence of mobile users and their traffic dynamics~\cite{salaht2020overview,yousefpour2019all}.
\par This paper pioneers a distributed service provisioning framework, aiming to dynamically harmonize for the first time the intricate interplay among QoS, service provisioning cost, and sustainability concerns within the edge-to-cloud ecosystem. The envisioned framework entails a reconfigurable edge-to-cloud computing infrastructure designed to balance the load of smart mobility services in response to dynamic spatio-temporal mobility patterns. The key contributions of this paper encompass: (i) Development of a novel and practical open-source framework\footnote{MERA framework, available at \url{https://github.com/DISC-Systems-Lab/Edge-Mobility-Cooptimization}} for dynamic provisioning (deploying and migrating) of smart mobility services. (ii) Formulation of an optimization problem addressing the QoS and mobility-aware service placement challenge. (iii) Applicability of an efficient collective decision making algorithm to tackle the problem, and (iv) Extensive evaluation based on real-world edge-to-cloud settings and traffic traces from Munich to verify the cost-effectiveness and scalability of the framework. 
\par This paper is organized as follows: The next section highlights our contribution to the existing literature. The proposed framework is introduced in Section~\ref{sec:spf}, followed by the formulation of the service placement problem in Section~\ref{sec:probform}. Then, a cooperative multi-agent approach for dynamic service placement is presented in Section~\ref{sec:sol}. Finally, Sections~\ref{sec:eval} and~\ref{sec:conc} provide experimental evaluations and conclusions, along with suggestions for future research.

\section{Related Work}
\noindent Service provisioning within edge-to-cloud network has garnered significant attention, leading to extensive exploration in both static and dynamic environments. Various optimization methods have been employed, including single~\cite{dai2018joint,rahbari2020task} and multi-objective~\cite{tran2018joint,kimovski2021mobility} optimization, linear programming~\cite{yousefpour2019fogplan,mouradian2019application}, Markov Decision Process (MDP)~\cite{kimovski2021mobility,kayal2019autonomic}, game theory~\cite{wang2020game,ouyang2018follow}, and fuzzy logic~\cite{tavousi2022fuzzy}, catering to diverse QoS criteria such as service delay~\cite{martin2020mobility,nezami2021decentralized,ouyang2018follow,matrouk2023mobility,mouradian2019application,djemai2021investigating}, resource efficiency~\cite{nezami2021decentralized}, energy consumption~\cite{djemai2020mobility,bahreini2021vecman,matrouk2023mobility,djemai2021investigating,djemai2019discrete}, economic cost~\cite{liu2023asynchronous,matrouk2023mobility,mouradian2019application}, and system utility~\cite{li2023cost,dai2018joint}. These formulations lay the groundwork for developing service placement algorithms falling into categories such as exact~\cite{happ2021joi,mann2020secure}, heuristic~\cite{chen2020computation,rahbari2020task,fan2022joint,tang2021toward}, meta-heuristic~\cite{djemai2019discrete,martin2020mobility,djemai2021investigating}, and machine learning-based~\cite{chen2020computation,liu2023asynchronous,zhu2020multiagent,sulimani2023reinforcement} solutions. 
Static placement~\cite{yousefpour2019fogplan,nezami2021decentralized,happ2021joi,mann2020secure,kayal2019autonomic,tavousi2022fuzzy,djemai2019discrete} becomes impractical over time in dynamic environments due to the impact of mobility on response time and infrastructure energy consumption. Especially in mobile scenarios that require stateful handling, migration becomes essential for optimizing efficiency. This section summarizes recent literature addressing the problem within a mobility-aware context, as detailed in Table~\ref{tab:lit-rev}.
\begin{table*}[ht]
    \centering
    \caption{Literature review summary}
    \label{tab:lit-rev}
    \footnotesize
    \begin{tabular}{p{1.5cm}p{2.2cm}p{2.8cm}p{2.2cm}p{2.5cm}p{4cm}}
        \toprule
        Criteria/ Study & Optimization Criteria & Proposed Approach & IoT Load & ICT Network & Traffic Network (mobility) \\
        \midrule
        \cite{liu2023asynchronous} & SU & DRL & Synthetic & Synthetic & - \\
        \cite{awada2023resource} & SD, RU & Bin-Packing & Alibaba cluster trace & - & Synthetic\\
        \cite{fan2022joint} & SD & Heuristic & Synthetic& - & Synthetic \\
        \cite{zhu2020multiagent} & SD & DRL & - &Synthetic & Rome (GPS trajectory) \\
        \cite{djemai2021investigating} & EC, SD & Greedy, Genetic algorithm & Synthetic & - & San Francisco (real trajectory) \\
        \cite{zhan2020deep} & SD, EC & DRL & Synthetic & - & - \\
        \cite{li2020read} & SR & Approximation & Synthetic & Synthetic & Melbourne (real trajectory) \\
        \cite{ouyang2018follow} &CD, MC & Approximation, Greedy & Synthetic &Synthetic& Helsinki (shortest paths) \\
        This work & EC, SD, RU, XC, CF & Collective learning & MAWI IoT trace & Munich edge-core-cloud network & Munich (shortest/default trajectory) \\
        \bottomrule
    \end{tabular}
        \vspace{0.5em} 
        \begin{spacing}{0.5} 
        \scriptsize
        \raggedright 
        \textbf{Abbreviation:} SU - System Utility, RU - Resource Utilization, SD - Service Delay, EC - Energy Consumption, XC - Execution Cost, CF - CO2 Footprint, SR - System Robustness, MC - Migration Cost.
        \end{spacing}
\end{table*}
Zhan \etal{}~\cite{zhan2020deep} introduce an MDP-based model to address service placement challenges in Vehicular Edge Computing (VEC) scenarios, focusing on reducing processing delay and system energy consumption. The authors leverage Deep Reinforcement Learning (DRL) methods to optimize this model. Awada \etal{}~\cite{awada2023resource} propose an integrated Mobile Edge Computing (MEC) framework for offloading computation-intensive applications to edge devices.
This framework consolidates edge resources from diverse locations into a unified pool, facilitating comprehensive monitoring from a single control plane. Fan \etal{}~\cite{fan2022joint} present a joint service offloading and resource allocation scheme for vehicles assisted MEC scenarios, aiming to minimize total processing delay through a two-stage heuristic algorithm that offers near-optimal solutions with low computational complexity.
Ouyang \etal{}~\cite{ouyang2018follow} introduce a centralized MDP approximation method and a distributed non-cooperative approach based on game theory for dynamic service placement to optimize communication and computing delay in mobile edge environments. The authors assume that wireless connections between users and edge nodes remain unchanged between placement rounds. Liu \etal{}~\cite{liu2023asynchronous} propose a vehicle-edge-cloud resource allocation framework aimed at maximizing system utility. This framework supports V2V offloading to enable vehicles with idle resources to engage in service deployment, and V2I offloading for load balancing. They employ a multi-agent MDP-based DRL algorithm to make optimal scheduling decisions.

\par The current research~\cite{liu2023asynchronous,awada2023resource,fan2022joint,li2020read,kimovski2021mobility,djemai2021investigating,chen2020computation,zhu2020multiagent,zhan2020deep,ouyang2018follow} on service placement typically focuses on a single or limited criteria. While service placement is inherently a trade-off problem, necessitating a comprehensive examination of multiple, often conflicting objectives. This research studies a holistic perspective encompassing sustainability, QoS, service cost, and resource efficiency within the heterogeneous edge-to-cloud infrastructure.
Application placement in a fog network is dynamic and autonomous, unlike the cloud, as fog nodes can join and leave the network freely. This variability in resource availability is compounded by fluctuations in demand for resources based on incoming request volume. Adjusting application placement at runtime to accommodate these changes is crucial, an aspect overlooked in current literature but addressed in this paper. 
This work supports adaptation to the V2I resources availability and service demands, ensuring flexibility and responsiveness to dynamic network conditions. Additionally, while most of the cited studies~\cite{liu2023asynchronous,awada2023resource,fan2022joint,djemai2021investigating} rely on central control plans or non-cooperative solutions~\cite{ouyang2018follow}, our approach stands out for its distributed cooperative nature, offering scalability, resilience, and flexibility.

\section{Service Placement Framework}\label{sec:spf}
\noindent This work defines a novel challenge in the realm of computing-mobility and tailors the system proposed in our prior work~\cite{nezami2021decentralized}, offering a comprehensive solution finely tuned to the mobility dynamics in the ecosystem. We present a Mobility-aware Edge-to-cloud Resource Allocation framework, MERA, that dynamically incorporates resource availability, mobility, and QoS considerations to address the challenges of service provisioning. The framework assesses resource utilization across the edge-to-cloud continuum by balancing factors such as load distribution, QoS fulfillment, economic cost, energy efficiency, and sustainability to fulfill the rigorous demands of smart mobility services, particularly those requiring low-latency and high-bandwidth capabilities. Improving QoS is particularly advantageous for applications such as real-time data analytics, augmented/virtual reality, industrial control with ultra-low latency, and big data streaming~\cite{yousefpour2019fogplan}. 
This section outlines the layered architecture of MERA, along with its request handling procedure, and practical considerations. Following this, the next section delves into problem formulation, considering both the individual and system-wide objectives of end-users, service providers, and ICT infrastructure providers.
\subsection{Motivational Application}
\noindent Based on the latest Cellular-Vehicle-to-Everything (C-V2X) roadmap of 5GAA\footnote{5G Automotive Association, available at \url{https://5gaa.org/about-us/} (accessed April 2024).}, the automotive and telecommunications industries are racing to enable technologies and applications such as HD Maps for fully driverless cars. These maps (e.g., Civil Maps\footnote{Civil Maps, available at \url{https://civilmaps.com/} (accessed April 2024).}) provide enhanced precision and accuracy, dynamically updated to reflect near real-time environmental conditions. Major players such as Google, Ford, BMW, Tesla, and General Motors are investing in HD Maps to advance autonomous vehicles. HD Maps aims to create a system where vehicles benefit from aggregated data on traffic patterns and construction zones, gathered by vehicles and sent over cellular networks. This enables commuters to contribute real-time incident reports to a precise crowd-sourced map accessible to all. Vehicles utilize this data via 4G LTE connectivity for automated mapping, overcoming challenges posed by frequent infrastructure changes, which traditional mapping struggles to address due to cost limitations in wireless data upload and processing.
\par In line with the development of HD Maps, ABI Research\footnote{ABI research, available at \url{https://www.abiresearch.com/press/augmented-reality-redefine-automotive-user-interfa/} (accessed April 2024).} forecasts a significant rise in Augmented Reality Head-Up Displays, with an estimated 15 million units shipped by 2025, including 11 million integrated into vehicles. This paper focuses on passenger-facing HD Maps' augmented reality technology, resembling the 2020 Mercedes-Benz GLE display~\cite{abdi2015vehicle}. Continuous data flows from vehicles to servers hosting AR services incur significant communication and processing costs, demanding substantial computational resources and low-latency processing~\cite{rao2012cloud}. This poses challenges to the readiness of self-driving cars, as highlighted by Civil Maps co-founder Sravan Puttagunta. We assume that all considered vehicles are equipped with onboard cameras, continuously uploading captured video/images to the edge-to-cloud infrastructure. Edge-to-cloud servers analyze data streams, providing updated maps to clients. The Sense-Process-Actuate model serves as the foundational framework for this application, starting with sensor data collection, conveying it as a continuous stream to higher-layer computing nodes for processing, and ending with command transmission to actuators~\cite{azizi2019priority,gupta2017ifogsim}.
\subsection{Distributed Architecture}
\noindent The system architecture of MERA within the V2I communication model is illustrated in Figure~\ref{fig:arch}. The bottom layer encompasses the traffic networks, serving as the infrastructure for delivering smart mobility services to end-users. This layer includes both congested and uncongested road networks, along with the mobility profiles of vehicles and supporting tools. Above the traffic networks, the data layer provides realism to the framework, incorporating real-world communication settings for LTE access/edge networks, core networks, cloud data centers, and IoT workloads.
\par Above the data layer lies a layered ICT network architecture designed to host smart mobility services originating from the transportation sector. This architecture utilizes networking and computation nodes across the edge-to-cloud hierarchy, with vehicles positioned at the bottom layer and cloud centers at the top layer, complemented by distributed fog servers positioned between them to facilitate efficient data processing and communication.
In the context of fog computing literature~\cite{hernandez2017implementing,gupta2017ifogsim}, fog nodes refer to computing or networking resources positioned between a data source and the central cloud. These nodes, which can include smartphones and routers, may be deployed at the cellular base station sites or data aggregation points such as a routers at the edge of the core network. 
Edge and core routers, switches, and cellular LTE base stations (referred to as access points) serve as the connectivity infrastructure within and between layers. Access points establish connections to the Internet via edge/core routers. Fog servers are interconnected with access points through edge routers, while the cloud is accessed through core routers~\cite{li2018end}. Vehicles are equipped with a cellular radio interface~\cite{malandrino2013content} to establish a connection and join the network upon entering the coverage range of a fixed access point. Access points, equipped with 4G LTE modules, connect to core/cloud nodes via a Wide Area Network, and they are interlinked through wireless backhaul.
\par At the apex of the architecture sits the service placement strategy, which orchestrates edge-to-cloud resources based on traffic patterns and service demands. This strategy plays a crucial role in optimizing edge-to-cloud resource utilization and ensuring the efficient delivery of smart mobility services within the edge-to-cloud infrastructure.
\subsection{Handling Requests and practicality}
\noindent Any vehicle requesting smart mobility services sends a service request, containing parameters outlined in Table~\ref{tab:notation}, along with its basic information (e.g., camera image, IP, mobility profile including velocity, timestamp, the orientation of the vehicle, and GPS coordinates) to its communicating fog server (which is the server connected to the vehicle's connecting access point) for processing. Each fog server is equipped with a software agent responsible for decision making regarding the placement of the requests received by the server itself. The offloading decision may be determined by considering the information about the agent's neighborhood and exchanged information with neighboring servers. This process ensures that placement objectives are achieved while adhering to a service level agreement (SLA). If the decision favors local execution, the request is processed by the node; otherwise, it is offloaded to a selected fog/cloud server. Upon acceptance, a container is instantiated at the designated server for request processing~\cite{nezami2023smotec}. The mapping of services to the available servers (i.e., service placement) is updated periodically as vehicles move around. Upon completion of service computation, host servers transmit the result to the nearest access point, which relays the response back to the vehicles.
\par The proposed framework is highly practical, operating without assumptions and requiring only minimal information about mobile nodes. For service placement, only the incoming resource demand from mobile nodes to fog nodes is required. Additionally, the framework may require the average transmission rate and propagation delay, approximated by the round-trip delay measured through a ping mechanism, between end-devices and their corresponding access point. The system has the flexibility to either own its fog resources or acquire them through rental arrangements with edge network providers such as AT\&T, Nokia, Verizon, or other edge resource owners~\cite{anglano2018profit,xu2017zenith,kim2019economics}. 
This work represents a rigor endeavor, grounded in real-world traffic and ICT architecture settings, specifically tailored for the vehicle-to-edge-to-cloud network paradigm. The authenticity of the data utilized in this study is ensured through selection from reputable sources, as detailed in Section~\ref{sec:eval}. 
\section{Mobility-aware Service Placement Problem Formulation}\label{sec:probform}
\noindent This section formally defines the mobility-aware service provisioning problem as an online task, presenting possible formulation of quadratic cost functions, acknowledging its NP-hard nature~\cite{bokhari1981mapping,nezami2021decentralized}. We explore formulations for both (i) System-wide objectives, emphasizing workload balance, incentivization of renewable energy sources, and infrastructure energy efficiency, and (ii) Local objectives, focusing on minimizing service provisioning costs and meeting QoS requirements. The solution to the service provisioning problem is a service placement plan ($\delta$) that contains placement decisions (i.e., binary variables), which place each service either on a fog server or on a cloud center. The binary variables ${x}_{ij}$ and ${x}_{ic}$ denote whether service i is placed on the fog node j or the cloud node c, respectively. $\overline{X}_{ij}$ denotes the initial configuration of i on j, which indicates whether j currently hosts the service. We consider a discrete time-slotted system model where time is divided into slots, and services are generated at the beginning of each slot. Then each time slot becomes a decision round. Hence, we denote both time slot and decision round as $\tau$ in seconds. Table~\ref{tab:notation} lists the notations used in the problem formulation.
\begin{figure}[!htbp]
\centering
\includegraphics[clip, trim=3.2cm 6.2cm 8.6cm 11.8cm, width=\columnwidth]{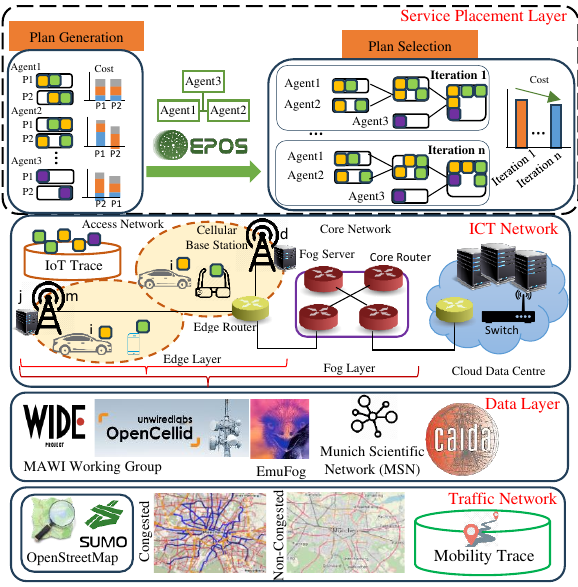}
\caption{Layered core components of MERA framework, consisting of: (1) Real-world traffic network, (2) Real-world data layer, (3) Vehicle-fog-cloud connectivity and computation infrastructure, and (4) Cooperative service placement orchestrating resource allocation.}
\label{fig:arch}
\end{figure}
\subsection{System-wide Objectives}
\noindent Due to the dynamic nature of vehicle mobility and the inherent unpredictability of the wireless medium, mobile IoT networks experience fluctuations in traffic load~\cite{bozorgchenani2021computation}. This research aims to assist policy makers and service providers in managing energy consumption and available edge-to-cloud capacity to handle incoming traffic effectively. Switching nodes off/on is a prevalent strategy for network management and resource optimization, providing various advantages such as improved energy efficiency, reduced operational costs, enhanced fault tolerance, and support for network scalability. However, this approach presents challenges in dynamic application placement, load balancing, and managing temporary network disruptions, especially in the time-varying environment of transport networks. Careful planning and monitoring are crucial for successful implementation.
This research proposes a load-balancing strategy involving selectively deactivating network nodes and redistributing the workload among the remaining active nodes. Furthermore, this research objective extends beyond merely achieving workload balance and reducing energy consumption; it also endeavors to advance renewable energy utilization through incentivization strategies. Energy harvesting from renewable resources such as wind and solar power holds significant promise for enhancing the efficiency and longevity of edge-to-cloud networks while also fostering sustainability from an environmental perspective~\cite{lee2015powering}.
\subsubsection{Workload Balance}
Leveraging fog nodes in transport environments enhances resource efficiency within edge-to-cloud networks and facilitates the execution of time-sensitive services. However, the individual objectives of fog agents may oppose both the broader system-wide goals and the interests of fog service providers or vehicles. Opting for the placement plans with the lowest costs can lead to the tragedy of the commons, where self-interested agents favoring the nearest edge servers may encounter elevated execution costs due to an uneven distribution of edge-to-cloud resource utilization. Effective load balancing, which mitigates the risk of bottlenecks, ensures a more robust and adaptable network infrastructure. By evenly distributing workload across the network, fog servers remain highly available to efficiently handle incoming requests, thereby reducing the likelihood of delayed responses to critical situations~\cite{khattak2019utilization,rahmani2015smart,gia2019exploiting}. The overarching objective is to minimize variance in server utilization (MIN-VAR), ensuring equitable load distribution across the network. Network nodes possess diverse capacities, necessitating the consideration of workload-to-capacity ratios to quantify node utilization. Equation~(\ref{eq:33}) illustrates the degree of workload balance across all nodes, measured by the variance in terms of CPU and RAM usage:
\begin{equation}
G = Min \sqrt{\frac{
\sum_{j=1}^{|{C}\cup{F}|} \left( \left( \frac{\zeta_{j}^{p}}{F_{j}^{\mathsf{p}}}-\overline{\frac{\zeta_{j}^{p}}{F_{j}^{\mathsf{p}}}} \right) ^{2} +
\left(\frac{\zeta_{j}^{m}}{F_{j}^{\mathsf{m}}}-\overline{\frac{\zeta_{j}^{m}}{F_{j}^{\mathsf{m}}}} \right) ^{2}
\right)}{2 \cdot (|F|+|C|)}},
\label{eq:33}\\
\end{equation}
\noindent where $\zeta_{j}^{m}$ and $\zeta_{j}^{p}$ denote the RAM and CPU load on node j and measured as $\sum_{i=1}^{|A|} \ L_{i}^{\mathsf{m}}\cdot x_{ij}$ and $\sum_{i=1}^{|A|} \ L_{i}^{\mathsf{p}}\cdot x_{ij}$, respectively.
\subsubsection{Energy Consumption versus Workload Balance}
The framework enables dynamic activation and deactivation of fog servers based on service demand, while also allowing for the adjustment of node participation levels in service hosting. To achieve this, an \emph{Infrastructure Planner} is defined, integrating traffic monitoring agents to observe incoming traffic rates~\cite{Guastella2023}. These agents, typically associated with Software Defined Networking (SDN) controllers such as OpenDayLight and ONOS, monitor aggregated IoT request traffic rates to fog nodes via northbound APIs~\cite{suarez2018sbar,luong2016traffic}. 
Using incoming traffic data and QoS considerations, the planner decides on activating or deactivating machines within the network, utilizing a resource configuration vector as a reference for fog servers. In response to increased demand, it may activate new machines in corresponding areas, while in regions with lower demand, it strategically powers down machines to conserve resources and minimize energy consumption. Consequently, the fog agents utilize the reference vector for conducting service placement and distributing the load on the active nodes.
\subsubsection{Renewable Energy Incentives}
Renewable energy can fully support the reliable operation of edge devices using a microgrid (solar-wind hybrid energy system)~\cite{li2018enabling}. This highlights the potential of renewable energy sources to sustain edge device functionality. It is assumed that edge-to-cloud servers and their associated networking devices are partially powered by clean energy, utilizing energy storage devices~\cite{ghiassi2014toward} to maximize renewable energy utilization. Nodes with a higher proportion of clean energy are proposed to be given priority in service deployment. To this end, (\ref{eq:27-1}) minimizes the squared Euclidean distance, Root Mean Square Error (RMSE), between network nodes' utilization and a target incentive vector based on the ratio of energy supplied from renewable sources.
\begin{equation}
G = \min \sqrt{\frac{\sum_{j=1}^{|F \cup C|}\left(\left(\frac{\zeta_{j}^{p}}{F_{j}^{\mathsf{p}}} - p_j^{\mathsf{r}}\right)^2 + \left(\frac{\zeta_{j}^{m}}{F_{j}^{\mathsf{m}}} - p_j^{\mathsf{r}}\right)^2\right)}{2 \cdot (|C| + |F|)}}
\label{eq:27-1}
\end{equation}
\subsection{Local Objective}\label{subsec:localcost}
\noindent The deployment of an application across the edge-to-cloud network can significantly influence its non-functional aspects such as operational costs and performance~\cite{islam2023optimal}. The overall cost of executing a typical service placement plan $\delta$ in edge-to-cloud infrastructure encompasses the following components: processing cost ($O^{\mathsf{P}}_{\delta}$), RAM usage cost ($O^{\mathsf{M}}_{\delta}$), storage usage cost ($O^{\mathsf{S}}_{\delta}$), container deployment cost ($O^{\mathsf{D}}_{\delta}$), communication cost ($O^{\mathsf{C}}_{\delta}$), energy consumption cost ($O^{\mathsf{E}}_{\delta}$), and carbon footprint cost ($O^{\mathsf{F}}_{\delta}$). As formulated in (\ref{eq:overalcost1}), this work combines these parameters with a penalty cost for service delay ($O^{\mathsf{V}}_{\delta}$), serving as a QoS measurement. Other expenses can also be identified such as security safeguards which are not the focus of this paper. 
\begin{equation}
L_{\delta} = O^{\mathsf{P}}_{\delta} + O^{\mathsf{M}}_{\delta} + O^{\mathsf{S}}_{\delta} + O_{\delta}^{\mathsf{D}} + O_{\delta}^{\mathsf{C}} + O_{\delta}^{\mathsf{E}} + O^{\mathsf{F}}_{\delta} + O_{\delta}^{\mathsf{V}}
\label{eq:overalcost1}
\end{equation}
\subsubsection{Cost of Processing and Memory Resources}
The processing cost in each node is variable and can differ from that of other nodes. Equation~(\ref{eq:3}) quantifies the processing cost associated with the placement plan $\delta$.
\begin{equation}
O^{\mathsf{P}}_{\delta} = 
\sum_{i=1}^{|A|}\sum_{j=1}^{|{F\bigcup C}|}x_{ij} \cdot z_{ij}  \cdot L_{i}^{\mathsf{p}} \cdot C_{j}^{\mathsf{p}} \cdot \tau
\label{eq:3}
\end{equation}
\par Similar to the processing cost, the memory cost for each node is specific to that node. Equations~(\ref{eq:4_1}) and~(\ref{eq:4_2}) calculate the RAM and storage cost for the plan $\delta$ respectively.
\begin{equation}
O^{\mathsf{S}}_{\delta} = \sum_{i=1}^{|A|}\sum_{j=1}^{|{F\bigcup C}|} \ x_{ij} \cdot (L_{i}^{\mathsf{s}} \cdot C_{j}^{\mathsf{s}}) \cdot  \tau
\label{eq:4_1}
\end{equation}
\begin{equation}
O^{\mathsf{M}}_{\delta} = \sum_{i=1}^{|A|}\sum_{j=1}^{|{F\bigcup C}|} \ x_{ij} \cdot (L_{i}^{\mathsf{m}} \cdot C_{j}^{\mathsf{m}}) \cdot \tau
\label{eq:4_2}
\end{equation}
\subsubsection{Cost of Service Delay}
IoT applications such as augmented reality are latency sensitive and have very rigid latency constraints in the order of tens of milliseconds~\cite{la2019enabling}, so that  a low latency is crucial to ensure an acceptable quality of experience. The service delay for an IoT service is defined as the time span between the moment an end-device sends its request and the moment it receives the response for that request. The binary variable $v_{ij}$ indicates whether the service delay $e_{ij}$ for service $i$ on server $j$ violates the SLA-defined delay threshold $h_{i}$.
\begin{align}
\scriptsize
v_{ij} & = \left\{
\begin{array}{rl}
0 & \text{if } e_{ij}<h_{i}\\
1 & \text{} otherwise
\end{array} \right.
\label{eq:8}
\end{align}
\normalsize
\par The delay of the service i includes three components as \emph{propagation delay}, \emph{transmission delay}, and \emph{waiting time} (processing delay plus queuing delay)~\cite{deng2016optimal,yousefpour2019fogplan}. 
Since vehicles may have different data generation rates, the time taken to execute a service depends on the data rate and the available computing capacity of the server to which it is assigned. 
In contrast to the computation time, which remains relatively constant, communication time varies over time due to fluctuations in latency along the communication path between mobile devices and the infrastructure. As shown in Figure.~\ref{fig:arch}, the mobility of vehicle i may result in dynamic changes in the communication links between the device and its service's host node j over time. These changes occur due to connection handoffs and handovers~\cite{puliafito2020mobfogsim}. Consequently, the dynamic nature of connections results in varying propagation delays and transmission rates for services within a given time interval.
To address this challenge in measuring service delay, we calculate those two parameters by considering the connecting access points along the vehicle's path during a specific time interval. The delay for a particular service is then computed using estimated values for propagation delay and transmission rate as follows:
\begin{equation}
\begin{aligned}
e_{ij} = &w_{ij}+\sum_{m=1}^{|M|}P_{im} \cdot (2 \cdot (d_{im}+d_{mj})\\
           &+(\frac{q_{i}}{b_{im}^{\mathsf{u}}}+\frac{a_{i}}{b_{im}^{\mathsf{d}}})+
            (\frac{q_{i}}{b_{mj}^{\mathsf{u}}}+\frac{a_{i}}{b_{mj}^{\mathsf{d}}})),
\end{aligned}
\label{eq:9}
\end{equation}
\noindent where $P_{im}$ signifies the probability of vehicle $i$ being within the coverage zone of access point $m$ during the time interval $\tau$ in a 2D area~\cite{saleem2021vehicle}. It is calculated as follows:
\begin{equation}
P_{im} = \frac{1}{\tau} \cdot (\frac{{l}_{im}}{s_{i}}-t_{im}-t^{\mathsf{w}}_{im}),
\label{eq:10}
\end{equation}
\noindent where ${l}_{im}$ represents the coverage area (Euclidean distance) between device $i$ and access point $m$ on street $x$ where $i$ is moving, $s_i$ denotes the speed of $i$, $t_{im}$ indicates the timestamp of the initial connection (registration time) of $i$ with $m$ before sending the service request, and $t^{\mathsf{w}}_{im}$ denotes the waiting time for $i$ to receive a reply after sending the service request to access point $m$.
\par To measure the waiting time ($w_{ij}$) of service i hosted on node j we adopt a multi-server M/M/C queuing model (i.e., Erlang-C)~\cite{gautam2012analysis} for fog servers. 
Fog node j has $n_{j}$ processing units, each with service rate $\mu_{j}$ (total processing capacity or service rate of node j is $F^{p}_{j} = n_{j} \cdot \mu_{j}$). For simplicity but without loss of generality, we assume that a cloud server similarly comprises of a set of homogeneous computing machines with similar configurations, such as CPU frequency.
As a result, the computation delay for service i hosted on either a fog or cloud server j is measured using (\ref{eq:11})~\cite{yousefpour2019fogplan,bolch2006queueing}.
\begin{equation}
w_{ij} =  
\frac{\mathrm{P}_{ij}^{\mathsf{Q}}}{F^{\mathsf{p}}_{j} \cdot f_{ij}-\zeta_{ij}^\mathsf{p}} +
\frac{1}{\mu_{j} \cdot f_{ij}},
\label{eq:11}
\end{equation}
\noindent where $\zeta_{ij}^\mathsf{p}$ denotes the arrival rate of instructions to node j for service i and measured as $\zeta_{ij}^\mathsf{p} = L_{i}^{\mathsf{p}} \cdot z_{ij} \cdot x_{ij}$, $f_{ij}$ denotes the fraction of processing units that service i deployed on node j can obtain and is measured by $f_{ij}=\frac{L_{i}^{\mathsf{p}} \cdot \ x_{ij}}{\sum_{i=1}^{|A|} L_{i}^{\mathsf{p}} \cdot x_{ij} }$. $\mathrm{P}^{\mathsf{Q}}_{ij}$, referred to as Erlang’s C formula, is measured as follows:
\begin{equation}
\mathrm{P}^{\mathsf{Q}}_{ij} = 
\frac{{(n_{j} \cdot \rho_{ij})}^{n_{j}}}{n_{j}!}.
\frac{\mathrm{P}^{\mathsf{0}}_{ij}}{1-\rho_{ij}},
\label{eq:12}
\end{equation}
\noindent where $\rho_{ij} = \frac{\zeta_{ij}^\mathsf{p}}{F^{\mathsf{p}}_{j} \cdot f_{ij}}$ and $\mathrm{P}^{\mathsf{0}}_{ij}$ is calculated as follows:
\begin{equation}
\mathrm{P}^{\mathsf{0}}_{ij} =  \left[{\sum_{c=0}^{n_{j}-1}\frac{(n_{j} \cdot \rho_{ij})^c}{c!} + \frac{(n_{j} \cdot \rho_{ij})^{n_{j}}}{n_{j}!}\left(\frac{1}{1-\rho_{ij}}\right)}\right]^{-1}
\label{eq:13}
\end{equation}
\par Finally, according to the QoS requirement, deadline violation is considered with respect to the percentage of requests from IoT services that their service delay exceed the delay threshold. 
We assume that desired quality for service\footnote{Amazon compute SLA, available at \url{https://aws.amazon.com/compute/sla} (accessed April 2024).} i is denoted by $\eta_{i}\,\epsilon\,(0,1)$ and the percentage of delay samples from services that exceed the delay threshold should be no more than $(1 - \eta_{i})$. We define $V_{ij}^{m}$ as the percentage of requests of service i hosted on node j that do not meet the delay requirement during the time connected to the access point m. Then $V_{i} = \sum_{m=1}^{|M|} V_{ij}^{m} \cdot P_{im}$ measures the total violation percentage for the service. To conclude, (\ref{eq:14}) calculates the cost of deadline violation for the placement plan $\delta$:  
\begin{equation}
O_{\delta}^{\mathsf{V}} = \sum_{i=1}^{|A|}\sum_{j=1}^{|{C}\bigcup{F}|}max (0 , V_{i}-(1-\eta_{i})) \cdot z_{ij} \cdot C_{i}^{\mathsf{v}} \cdot x_{ij} \cdot \tau 
\label{eq:14}\\
\end{equation}
\subsubsection{Cost of Service Deployment}
Given the movements of end users, it is crucial for services to dynamically migrate across multiple nodes to maintain service performance and minimize user-perceived latency. However, frequent migration significantly increases operational costs. To address this trade-off, we investigate deployment cost, which measures the communication cost of service deployment from cloud nodes to fog nodes. Clients of service providers develop and upload new services to public cloud storage, and these services are then downloaded onto target nodes upon deployment. When the demand for a deployed service decreases, the host node may release the service to conserve space. If a fog node receives requests for a service not locally hosted, it must download and deploy the service locally. Notably, a cloud center has virtually unlimited storage space and can host services for an extended duration. As a result, the communication cost for service deployment on the cloud is omitted and (\ref{eq:5}) measures the cost of service deployment.
\begin{equation}
O_{\delta}^{\mathsf{D}} = 
\sum_{i=1}^{|A|}\sum_{j=1}^{|{F}|}x_{ij} \cdot (1-\overline{x}_{ij})\cdot L_{i}^{\mathsf{s}} \cdot C_{jc}^{\mathsf{c}}, 
\label{eq:5}\\
\end{equation}
\noindent where $C_{jc}^{\mathsf{c}}$ is the unit cost of communication between the fog node j and the cloud node c.
\subsubsection{Cost of Communication}
This cost encompasses data upload/download overhead, including service requests and responses transmission. Services may be hosted on nodes different from those directly connected, requiring data transfer from the connecting access point to the host node. In addition, vehicle mobility may lead to traversing multiple access points during service offloading/processing. These factors contribute to fluctuating communication costs over defined time periods. 
In the illustrated example in Figure~\ref{fig:arch}, the service requested by vehicle i is intended to be placed on fog node j. At the start of the time interval ($t_{0}$), i is linked to j through m, and by the end of the interval ($t_{0}+\tau$), it is connected to j through access point d. A relay mechanism, commonly referred to as ``handoff'' or ``handover,'' allows a vehicle to obtain the service result from a neighboring access point if receiving it from the original access point is impractical.
In other words, access points communicate with each other, and the service is relayed through the backhaul network~\cite{zhang2017mobile,janevski2003traffic}. 
\par Intermittent connectivity due to vehicle movement poses a significant risk to successful data transmission. Therefore, ensuring reliable link connectivity is crucial for the success of computation offloading. This reliability can be measured by the duration of link connections. Accordingly, (\ref{eq:6}) measures the relevant communication cost.
\begin{equation}
O_{\delta}^{\mathsf{C}} = \sum_{i=1}^{|A|} \sum_{j=1}^{|{C}\bigcup{F}|} \sum_{m=1}^{|M|} x_{ij} \cdot (q_{i}+a_{i}) \cdot z_{ij} \cdot C_{mj}^{\mathsf{c}} \cdot {\tau}_{im},
\label{eq:6}
\end{equation}
\noindent where the connectivity time ${\tau}_{im}$  represents the time during which vehicle i remains connected to access point m as ${\tau}_{im} = P_{im} \cdot \tau$ during the time interval $\tau$.
\subsubsection{Cost of Energy Consumption}\label{subsec:localcost-pc}
The energy consumption of an IoT network comprises both static and dynamic components. The static part pertains to energy usage when resources are idle~\cite{kurpicz2016much}, while the dynamic component is determined by the current workload on active virtual machines and containers~\cite{ahvar2019estimating,salaht2020overview}. As the static part remains unaffected by service placement policies, we focus solely on the dynamic component in our model. A significant portion of this component stems 
from computationally intensive requests~\cite{ismail2021escove}, primarily attributed to networking and server machines.
\begin{equation}
O_{\delta}^{\mathsf{E}} = \tau \cdot (P^{\mathsf{n}}_{\delta} + P_{\delta}^{\mathsf{s}})
\label{eq:18}
\end{equation}
\par The dynamic power consumption of physical machines primarily depends on CPU utilization and can be modeled as a linear function~\cite{wiesner2021leaf,lee2012energy,heinrich2017predicting} (refer to (\ref{eq:22})). Our fog servers, resembling nano data centers, are composed of single physical machines without additional ICT equipment such as fans and links~\cite{ahvar2019estimating}. To incorporate the energy consumption of additional ICT equipment, we utilize the Power Usage Effectiveness (PUE), denoted by $\theta_{c}$, which signifies the ratio between total facility power consumption and IT equipment power consumption.
\begin{equation}
\begin{aligned}
P_{\delta}^{s} \triangleq &\sum_{j=1}^{|F|}(P_{j}^{\mathsf{a}}-P_{j}^{\mathsf{i}}) \cdot u_{j} \cdot
\left[
(1-p_{j}^{\mathsf{r}}) \cdot C^{\mathsf{n}}+p_{j}^{\mathsf{r}} \cdot C^{\mathsf{r}}
\right]\\
&\sum_{c=1}^{|C|}\theta_{c} \cdot (P_{c}^{\mathsf{a}}-P_{c}^{\mathsf{i}}) \cdot u_{c} \cdot
\left[
(1-p_{c}^{\mathsf{r}}) \cdot C^{\mathsf{n}}+p_{c}^{\mathsf{r}} \cdot C^{\mathsf{r}}
\right],
\label{eq:22}
\end{aligned}
\end{equation}
\noindent where the parameter $(P_{j}^{\mathsf{a}}-P_{j}^{\mathsf{i}})/F_{j}^{\mathsf{p}}$ determines the incremental power consumption per unit load on the servers and $u_{j}=\frac{\zeta_{j}^{p}}{F_{j}^{\mathsf{p}}}$.
To compute the dynamic energy consumption of the telecommunication network linking physical servers with vehicles we consider three main types of networking devices: edge routers, core routers, and switches.
When a networking equipment carries traffic, it consumes load-dependent energy for packet processing and also for storing and forwarding the payload~\cite{vishwanath2014modeling,sivaraman2011profiling}. Consequently, if the router j (the router connecting the fog server j to the network) receives $z_{ij}$ requests for service i during the time period $\tau$, it processes $z_{ij} \cdot (a_{i}+q_{i})$ bytes in such interval. Note that a cloud center is linked to the fog infrastructure through an edge router, whereas fog servers can be connected via either core routers or edge routers and access points. Additionally, mobile devices may connect to various access points within a given interval. Putting it all together, (\ref{eq:25}) presents the networking power consumption of edge-to-cloud infrastructure as a result of the placement plan $\delta$. The first segment in (\ref{eq:25}) calculates the power consumption of networking devices (edge and core routers) connecting host nodes to the network. The subsequent part quantifies this metric for the switch within cloud centers, while the third segment assesses the energy consumed by access points connecting mobile devices to the network.
\begin{equation}
\begin{aligned}
P^{n}_{\delta} = &\sum_{i=1}^{|A|} \sum_{j=1}^{|F|} x_{ij} \cdot z_{ij} \cdot p_{j}^{\mathsf{f}} \cdot (a_{i}+q_{i}) \cdot \\
&\left[
(1-p_{j}^{\mathsf{r}}) \cdot C^{\mathsf{n}}+p_{j}^{\mathsf{r}} \cdot C^{\mathsf{r}}
\right]+\\
&\sum_{i=1}^{|A|} \sum_{c=1}^{|C|} \theta_{c} \cdot x_{ic} \cdot z_{ic} \cdot p^{\mathsf{f}}_{\mathsf{c}} \cdot (a_{i}+q_{i}) \cdot \\
&\left[
(1-p_{c}^{\mathsf{r}}) \cdot C^{\mathsf{n}}+p_{c}^{\mathsf{r}} \cdot C^{\mathsf{r}}
\right]+\\
&\sum_{i=1}^{|A|} \sum_{j=1}^{|F \bigcup C|} \sum_{m=1}^{|M|} x_{ij} \cdot P_{im} \cdot z_{ij} \cdot p_{m}^{\mathsf{f}} \cdot (a_{i}+q_{i}) \cdot \\
&\left[
(1-p_{m}^{\mathsf{r}}) \cdot C^{\mathsf{n}}+p_{m}^{\mathsf{r}} \cdot C^{\mathsf{r}}
\right]
\label{eq:25}
\end{aligned}
\end{equation}
\subsubsection{Cost of Carbon Footprint}
According to the Shift Project report\footnote{Environmental impact of digital: 5-year trends and 5G governance, available at \url{https://theshiftproject.org/article/impact-environnemental-du-numerique-5g-nouvelle-etude-du-shift/} (accessed April 2024)}, carbon emission from information technology infrastructure and data servers supporting cloud computing now surpasses those from pre-Covid air travel. The carbon footprint is a critical consideration in server equipment deployment, and this study incorporates its cost into the optimization process, as shown in (\ref{eq:27}). Here, $R_{c}$ represents the average carbon emission rate for electricity per kilogram per kilowatt-hour~\cite{hussain2019fog}. The calculation only accounts for power consumption from non-renewable sources ($P_{\delta}^{E}$), following the U.S. Energy Information Administration's classification of clean energy sources as carbon neutral.
\begin{equation}
O^{\mathsf{F}}_{\delta} = C^{f} \cdot R_{c} \cdot P_{\delta}^{\mathsf{E}} \cdot \tau,
\label{eq:27}
\end{equation}
\subsection{Constraints}
\noindent Resource utilization of fog nodes and cloud servers must not exceed their capacity, as formulated by:
\begin{equation}
\sum_{i=1}^{|A|} \ L_{i}^{\mathsf{s}} \cdot x_{ij}<F_{j}^{\mathsf{s}}\text{, and} \sum_{i=1}^{|A|} \ L_{i}^{\mathsf{m}} \cdot x_{ij}<F_{j}^{\mathsf{m}} \:,\forall j\,\epsilon\,F
\label{eq:28}
\end{equation}
\begin{equation}
\sum_{i=1}^{|A|} L_{i}^{\mathsf{s}}  \cdot x_{ic}<F_{c}^{\mathsf{s}}\text{, and} \sum_{i=1}^{|A|} L_{i}^{\mathsf{m}}  \cdot x_{ic}<F_{c}^{\mathsf{m}}\:,\forall c\,\epsilon\,C
\label{eq:29}
\end{equation}
In addition, stability constraints of the queues for the services on fog nodes and cloud servers imply:
\begin{equation}
\zeta_{ij}^\mathsf{p}<F_{j}^{\mathsf{p}} \cdot f_{ij}\text{, and}\
\zeta_{ic}^\mathsf{p}<F_{c}^{\mathsf{p}} \cdot f_{ic} \:\forall j\,\epsilon\,F,\forall c\,\epsilon\,C , \forall i\,\epsilon\,A 
\label{eq:30}\\
\end{equation}
Finally, the placement of services is constrained so that each service must be hosted at most on one computational resource. Formally:
\begin{equation}
0\leq \sum_{i=1}^{|A|}\sum_{j=1}^{|{C}\bigcup{F}|}x_{ij}\leq |A|
\label{eq:32}
\end{equation}
\subsection{Final Optimization Formulation}
\noindent To address the overall problem, all eight local cost functions are considered initially; however, in certain scenarios, some costs can be omitted if needed, while others may play a dominant role in the overall cost summation. Note that in this optimization problem, we only consider the costs of fog-to-fog and fog-to-cloud communication. In other words, the cost of communication between IoT and fog is not considered in the optimization problem, since this is usually outside the control of service providers.
Achieving both system-wide and local objectives requires coordination among agents and cannot be achieved solely through local optimization. Agents must select placement plans that meet multiple complex criteria, including minimizing quadratic cost functions. The next section introduces our proposed placement algorithm that autonomously manages these diverse objectives, leveraging agents' local parameters to effectively navigate trade-offs.

\section{Cooperative Service Placement Algorithm}\label{sec:sol}
\noindent This section illustrates how collaboration among fog agents can enhance both system-wide and individual benefits. It outlines a distributed service placement algorithm, addressing the challenges of the problem in periodic intervals, avoiding centralized servers' bottlenecks, and implementing a strategy for machine activation and deactivation to ensure scalability against bursty resource usage patterns. As shown in Figure.~\ref{fig:arch}, fog agents collaborate to select one service placement plan for their received requests in the two steps of plan generation and plan selection, aiming to meet both local (minimization of service provisioning cost) and global (e.g., minimization of utilization variance) objectives. This study significantly advances the groundwork laid by prior research~\cite{nezami2021decentralized} by introducing a refined methodology. It incorporates eight distinct cost components during the plan generation phase and seamlessly integrates them with three overarching system-wide objectives in the subsequent plan selection process.
\par Initially, each fog agent autonomously generates multiple mappings (i.e., possible service placement plans with different costs). In this step, agents prioritize minimizing the service placement cost of their own received requests to enhance QoS and keep the service provisioning cost low. The possible plans encode selected hosts using a binary vector and resource utilization with a vector of real values~\cite{nezami2021decentralized}. The utilization vector reflects resource usage by representing the ratio of the assigned load to the capacity for each host node. This approach allows us to consider the heterogeneity in the capabilities of these nodes, providing a more nuanced representation of their individual capacities.
As a result of the plan generation step, each agent calculates a set of potential alternative plans, each associated with the respective local cost. 

\par To address the system-wide objectives, subsequently, all agents engage in collaborative decision-making to select one plan among their possible plans based on both local and global objectives. This research leverages EPOS (Economic Planning and Optimized Selections)~\cite{pournaras2018decentralized,Pournaras2020}, a decentralized multi-agent system designed to tackle complex multi-objective discrete-choice combinatorial problems via a collective learning approach. It expands the scope of EPOS applicability by bridging computing and mobility, allowing for an examination of their interplay. Agents autonomously organize into a tree overlay topology to structure their interactions and facilitate information exchange~\cite{pournaras2018decentralized,Pournaras2020}, thereby enabling cooperative optimization. The optimization unfolds through a series of successive learning iterations, encompassing two phases: plan aggregation occurs in a bottom-up (leaves to root) fashion, while feedback propagation follows a top-down (root to leaves) approach within this structure. In each iteration, agents refine their chosen plans by combining the two objectives in a weighted sum of costs, as depicted in (\ref{eq:comcost}). This weighted summation facilitates making trade-offs and supports multiple levels of QoS.
\begin{equation}
\lambda \cdot L^t + (1-\lambda) \cdot G^t,
\label{eq:comcost}
\end{equation}
\noindent where $\lambda \epsilon [0, 1]$. The higher the value of the weight, the stronger the preference towards minimizing the corresponding objective. The cost functions take as an argument the global plan ($g$) at the iteration $t-1$, which is the sum of all utilization plans of the agents in the network. The global cost function and the local cost function are formulated as follows:
\begin{equation}
 G^t = \sigma (g^t),\ L^t = min \frac{1}{|F|}\sum_{j=1}^{|F|} l(\delta _j^t), 
\label{eq:overalcost}
\end{equation}
\noindent where $G^t, L^t \epsilon \mathbb{R}$ and $l(.)$ extract the local cost of the selected plan $\delta$ of the agent j at iteration t and $|F|$ represents the number of fog agents engaged in the decision making.
\par MERA employs a local plan generation approach, which iterates for a specified number of plans. Consequently, its computational complexity is contingent upon on the number of iterations conducted within EPOS, with the critical path's complexity linked to the network size. In EPOS, the tree height dictates the computational load, which scales logarithmically with the number of agents, denoted as $log |F|$. For each agent, the complexity is $O(|\delta| \cdot t)$. As a result, the overall complexity for all services is $O(|\delta|\cdot t \cdot\log |F|)$. In the subsequent section, we conduct experimental evaluations of the algorithm using real-world data.
\section{Evaluation}\label{sec:eval}
\noindent Six scenarios are utilized to assess the effectiveness of the proposed framework. These scenarios encompass three service placement strategies, including \emph{MERA}, alongside two widely recognized approaches: \emph{Baseline} and \emph{Greedy}. The evaluations are conducted across two distinct types of mobility profiles: \emph{default} and \emph{optimized} routes for vehicles.
In the Baseline approach, each fog server evaluates its own resource availability. If feasible, local deployment is chosen; otherwise, requests are forwarded to the cloud. The Greedy algorithm, referring to the widely-used First Fit approach in edge and cloud computing~\cite{brent1989efficient,skarlat2017,yousefpour2019fogplan}, aims to efficiently allocate resources while balancing optimization with computational efficiency. Upon receiving a request, a fog server assesses nearby resources with adequate capacity for the service, computes the execution cost on each server, and sorts them based on ascending cost. It then selects the most suitable option from the beginning of the sorted list, iterating this process for subsequent requests.
The Baseline approach focuses on offloading services solely to directly connected edge servers, aiming to minimize costs by utilizing local resources and a local perspective.
Conversely, the Greedy approach pursues optimal solutions locally, leveraging a global view of the system.
\par
\begin{figure} 
    \centering
    \begin{subfigure}{0.23\textwidth} 
        \fbox{\includegraphics[clip, trim=0.2cm 0.5cm 0.2cm 0.1cm, width=4cm]{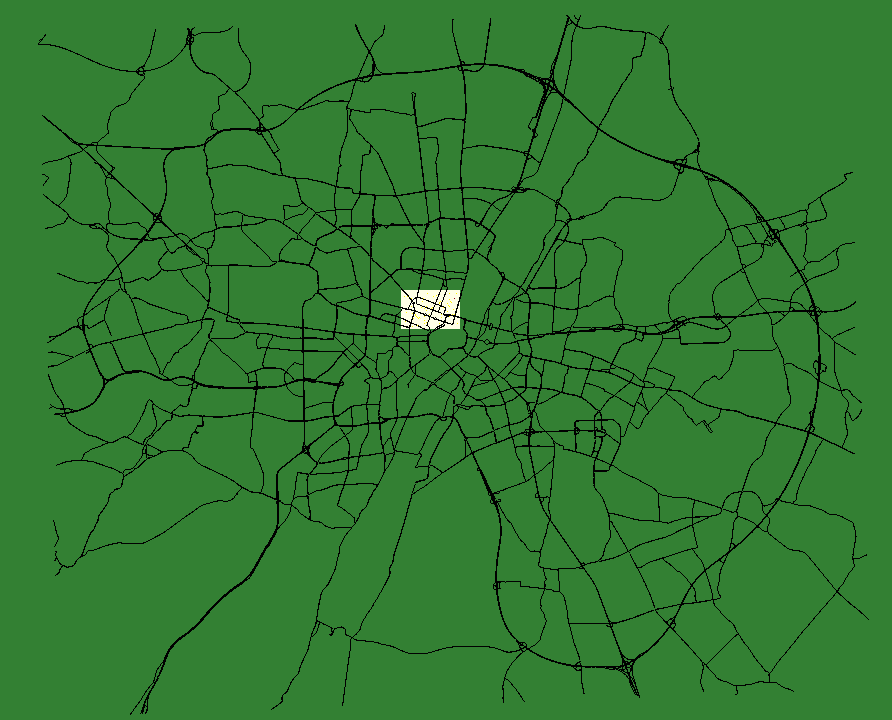}}
    \end{subfigure}%
    \hspace{\fill} 
    \begin{subfigure}{0.25\textwidth} 
        \fbox{\includegraphics[clip, trim=1.2cm 1.2cm 1.8cm 0.8cm, width=4cm, height=3.15cm]{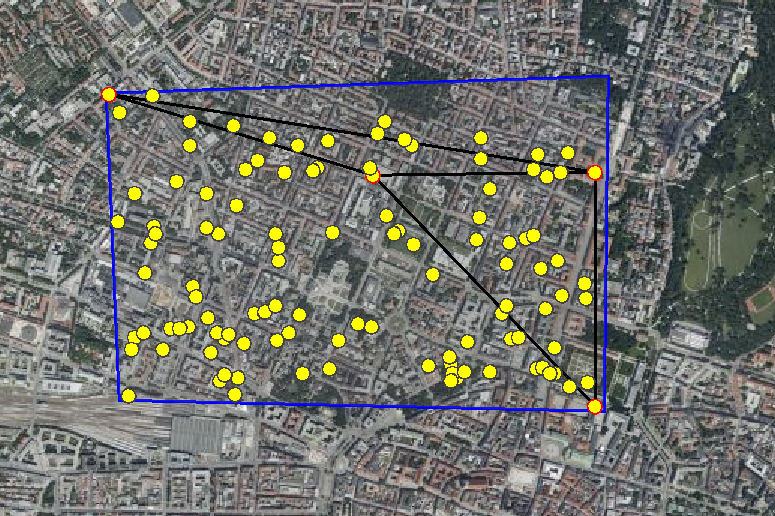}}
    \end{subfigure}%
    \caption{Rectangular test area in Munich city center, core routers connected with black lines and LTE access points within the test area are highlighted.}
    \label{fig:Testarea}
\end{figure}

\begin{figure}
		\includegraphics[clip, trim=0cm 0.8cm 0.2cm 0.15cm, width=9cm]{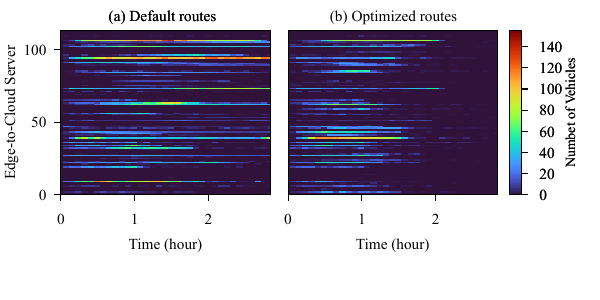}
        \caption{Traffic distribution across the edge-to-cloud network: the number of connected vehicles to each access point, along with their presence time, is unbalanced and higher in default routes compared to optimized routes.}
		\label{fig:vehtraffic}
  \vspace{-10pt}
\end{figure}

 The experiments are conducted within Munich, Germany. An area of $7000 {m}^{2}$ of the city center is selected as the test area, shown in Figure.~\ref{fig:Testarea}. 
The traffic distribution pattern is derived from a real-world traffic trace representing the vehicle journeys within Munich city and the Sumo traffic simulator\footnote{SUMO, available at \url{https://sumo.dlr.de/docs/index.html} (accessed April 2024).}, employing two routing algorithms: (i) The Dijkstra algorithm, which determines the shortest paths for vehicles based on travel time and distance (default routes), and (ii) The duaIterate algorithm\footnote{Duarouter, available at \url{https://sumo.dlr.de/docs/duarouter.html} (accessed April 2024).}, which iteratively generates and refines the routes of vehicles to enhance traffic flow and alleviate congestion (optimized routes). Figure~\ref{fig:vehtraffic} shows the number of vehicles within the area and connected to the edge-to-cloud servers using the aforementioned routing algorithms over a three-hour period.

\normalsize

\subsection{Simulation Setting}
\begin{figure}[!htbp]
	\includegraphics[clip, trim=5.5cm 11.2cm 5.4cm 11.5cm,width=9cm, height=4cm]{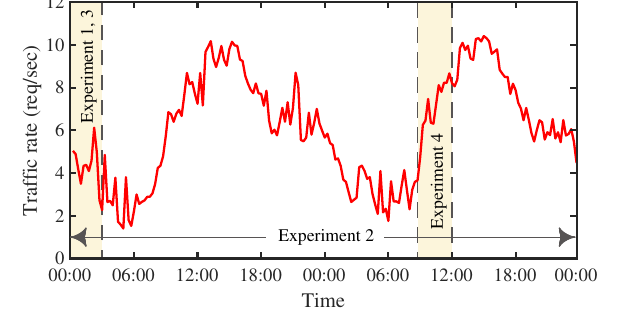}
	\caption{48 hours (192 15-minute IoT profiles) of MAWI trace data (April 12-13, 2017) serve as the input workload, delineating the experimental time intervals}
		\label{fig:iottraffic}
\end{figure}

\noindent The incoming workload to the experiments originates form two sources (i) Vehicles' traffic, indicated in Figure.~\ref{fig:vehtraffic}, referred to as mobility profiles and (ii) IoT service requests, indicated in Figure.~\ref{fig:iottraffic}, referred to as IoT profiles. 
The IoT service requests come from the real-world traffic traces of Measurement and Analysis on the WIDE Internet (MAWI) traffic repository\footnote{MAWI working group traffic archive, available at \url{http://mawi.wide.ad.jp/mawi} (accessed April 2024).}. The MAWI archive, derived from the WIDE Internet backbone connecting Japanese academic and research institutions to the Internet, serves as a realistic dataset for modeling incoming IoT traffic rates to fog nodes~\cite{yousefpour2019fogplan}. Capturing daily traces at the transit link of WIDE to its upstream ISP, the archive includes packets covering common IoT protocols such as MQTT, COAP, mDNS, and AMQP that utilize TCP and UDP for message transmission.
Figure~\ref{fig:iottraffic} illustrates the average incoming traffic rate ($z_{ij}$) to the fog nodes per service within a 2-day time span. This traffic is subsequently distributed across the vehicles in the test area using a uniform distribution. Table~\ref{tab:testparams} outlines additional characteristics of a mobile augmented reality application continuously running on each vehicle for autonomous driving~\cite{wintersberger2019fostering,feng2018augmented}. 
\begin{center}
\tablefirsthead{%
\hline
{Parameter}&{Value}&{Parameter}&{Value}\\
\hline}
\tablehead{\hline
{Parameter}&{Value}&{Parameter}&{Value}\\
\hline}
\tabletail{%
\hline
\multicolumn{2}{r}{\small\sl continued on next column}\\
\hline}
\tablelasttail{\hline}
\topcaption{Experimentation parameters~\cite{vishwanath2013energy,yousefpour2019fogplan}}\label{tab:testparams}
\tiny
\begin{supertabular}{|p{0.7cm}p{2.8cm}|p{0.7cm}p{2.8cm}|}
\hline
	Test Area&$2101.98*3313.97m^{2}$&$L_{i}^{\mathsf{m}}$&U(2-400) MB\\
	$|C|$&1&$L_{i}^{\mathsf{p}}$&U(50,200) MI per req\\
	$|F|$&114&$L_{i}^{\mathsf{s}}$&U(50-500) MB\\
	$|S|$&1&$h_{i}$&10 ms\\
	$|M|$&109&$q_{i}$&U(10-26) KB\\
	$|A|$&{9788,10112}&$a_{i}$&U(10-20) Byte\\ 
 
    $|N|$&109&$|\delta|$&20\\
        
	$s_{i}$ & $\leq 120 km/h$&$\theta_{c}$&1.2\footnote{Cloud computing, server utilization, \& the environment, available at \url{https://aws.amazon.com/blogs/aws/cloud-computing-server-utilization-the-environment/} (accessed April 2024).}\\
 
    $\tau$&300 seconds&$p^{\mathsf{r}}_{c}$&0.85\footnote{International Energy Agency, available at \url{https://www.iea.org/energy-system/buildings/data-centres-and-data-transmission-networks} (accessed April 2024).}\\
    
     $\lambda$ & $\{0.05n \mid n \in \mathbb{N}_0, 0 \leq 0.05n \leq 1.0\}$ & $p^{\mathsf{r}}_{j}$ & $\text{Beta}(2, 0.5)$ \\
       
 
        $C_{j}^{\mathsf{p}}$&edge 0.6, otherwise 0.2 USD per 1M req\footnote{AWS pricing, available at \url{https://aws.amazon.com/lambda/pricing/} (accessed April 2024).}&$R_{c}$&380 g CO2 eq/kWh\footnote{Current emission in Germany, available at \url{https://www.nowtricity.com/country/germany/} (accessed April 2024).}\\
        
        $C_{j}^{\mathsf{m}}$&$21*10^{-10}/128$ USD per MB per ms\footnote{AWS lambda pricing for a serverless application, available at \url{https://www.clickittech.com/devops/aws-lambda-pricing/}(accessed April 2024).}&$b^{\mathsf{d}}_{jj'}$&[LTE: 0.072, Edge/Core: 10/100, Cloud: 100] Gbps\\ 
        
        $C_{j}^{\mathsf{s}}$&U(0.021,0.023) USD per GB per month\footnote{AWS S3, available at \url{https://aws.amazon.com/s3/pricing/}(accessed April 2024).}&$b^{\mathsf{u}}_{jj'}$&[LTE: 0.012, Edge/Core: 1/10, Cloud: 100] Gbps\\
        
        $C_{jj'}^{\mathsf{c}}$&U(0.01,0.03) USD per GB (intra edge),
        U(0.03, 0.06) \$ per GB (intra core or core-edge),
        U(0.06, 0.09) \$ per GB (cloud communication)\footnote{The guide to AWS data transfer pricing and saving, available at \url{https://www.cloudbolt.io/guide-to-aws-cost-optimization/aws-data-transfer-pricing/} (accessed April 2024).}&$d_{ic}$&U(15, 35) ms\\
        
        $C^{\mathsf{v}}_{i}$&[$\eta_{i} \kern-2pt<\kern-2pt 95.0\%$: 100\%, 
        $95.0\% \kern-2pt < \kern-2pt \eta_{i} \kern-2pt<\kern-2pt 99.0\%$: 30\%,
        $\eta_{i} \kern-1pt<\kern-1pt 99.5\%$: 10\%] Service Credit Percentage\footnote{AWS SLA, available at \url{https://aws.amazon.com/compute/sla/} (accessed April 2024).}&$C_{c}$&17.27 \euro{} per tonnes\footnote{Carbon taxes in Europe, available at \url{https://taxfoundation.org/carbon-taxes-in-europe-2022/} (accessed April 2024).}\\
        
        $C^{\mathsf{n}}$&0.905 USD per KWh\footnote{Germany electricity prices, available at \url{https://www.globalpetrolprices.com/Germany/electricity_prices/} (accessed April 2024).}&$C^{\mathsf{r}}$&294 \euro{} per MWh\footnote{S\&P global commodity insights, available at \url{https://www.spglobal.com/commodityinsights/en/market-insights/latest-news} (accessed April 2024).}\\        
\hline
\end{supertabular}
\end{center}

The IoT workload is updated every 15 minutes based on the IoT trace data. Each 15-minute profile is divided into three 5-minute runs. Additionally, the number of vehicles in the test area is updated every 5 minutes. At the beginning of each 5-minute interval, ($\tau$), the placement algorithms are executed, and the resulting placements are assumed to remain constant throughout the interval.
\subsubsection{Edge-to-Cloud Network}
The experiments are conducted on the world's largest Open Database of Cell Towers (OpenCellID\footnote{OpenCelliD, available at \url{https://opencellid.org} (accessed April 2024).} dataset) that provides the locations of real-world cellular base stations. A total of 109 LTE Base Transceiver Stations situated within the Munich test area, as illustrated in Figure~\ref{fig:Testarea}b. 
In the densely populated urban core of downtown Munich, each cell site is assumed to connect adjacent vehicles to the network within a coverage range of 700 meters, operating in the 2.6 GHz frequency band~\cite{patlayenko2021comparison}.
The network includes one cloud center located within a 10-hop distance from access points, with propagation delay U(15, 35) ms~\cite{yousefpour2019fogplan}, and each cloud center requiring a Cisco Ethernet switch for operation~\cite{ahvar2019estimating}.
The number and location of core routers within the test area is determined using the Macroscopic Internet Topology Caida dataset\footnote{Macroscopic Internet topology, available at \url{https://www.caida.org/data/internet-topology-data-kit/release-2019-04.xml} (accessed April 2024).}, EmuFog emulator~\cite{mayer2017emufog}, and the Munich Scientific Network (MWN)\footnote{The Munich scientific network, available at \url{https://www.lrz.de/services/netz/} (accessed April 2024).}. Tables~\ref{tab:testparams} and \ref{tab:routers} outline the specification of the networking devices and their capacity.       
\begin{table*}[t]
\setlength\tabcolsep{2pt}
\scriptsize\centering
\caption {Hardware specification and power consumption of the servers used in the evaluation}
\label{tab:machines}
\smallskip
\begin{tabular*}{16cm}{p{5cm}p{1.6cm}p{2.7cm}p{1.7cm}p{2cm}p{2cm}}
\hline
Machine model&Memory (GB)&Performance (GFLOPS)&Storage (GB)&Idle power (Watt)&Max power (Watt)\\
\hline
Xeon Gold 6140 36cores 2.3GHz&192&864&120&52.4&343\\
Xeon Gold 6136 24cores 3.0GHz&196&806.4&292&131&432\\
Xeon Platinum 8180 56cores 2.5GHz&192&1523.2&480&48&385\\
Xeon Platinum 8280 56cores 2.7GHz&192&1612.8&240&64.2&435\\
Xeon Platinum 8380HL 112cores 2.9GHz&384&1792&480&44.6&502\\
Cloud Xeon E5-2680 10cores 2.80 GHz&768&112000MIPS&500&57&115\\
\hline
\end{tabular*}
\end{table*}

\begin{table*}[t]
\setlength\tabcolsep{2pt}
\scriptsize\centering
\caption {Networking equipment and their characteristics~\cite{vishwanath2013energy,vishwanath2014modeling, sivaraman2011profiling}}
\label{tab:routers}
\smallskip
\begin{tabular*}{16cm}{p{4.3cm}p{1.2cm}p{1.2cm}p{2.2cm}p{2.0cm}p{1.8cm}p{1.5cm}}
\hline
    Device type&Idle power (Watt)&Max power (Watt)&Download max traffic capacity (Gbps)&Upload max traffic capacity (Gbps)&Download energy (Nj/bit)&Upload energy (Nj/bit)\\
\hline
   3-sector 2×2 MIMO LTE base station&333&528&0.072&0.012&82820&12400\\
   Cisco edge router 7609&4095&4550&560&560&37&37\\
   Cisco core router CRS-3&11070&12300&4480&4480&12.6&12.6\\
   Cisco Ethernet switch Catalyst 6509&1589&1766&256&256&31.7&31.7\\
\hline
\end{tabular*}
\end{table*}
\par The machines detailed in Table~\ref{tab:machines} exhibit comparable characteristics to the c5.metal instances introduced by Amazon in 2019\footnote{Amazon EC2 C5 instances, available at \url{https://aws.amazon.com/blogs/aws/now-available-new-c5-instance-sizes-and-bare-metal-instances/} (accessed April 2024).}, serving as our fog and cloud servers. Furthermore, it is assumed that the storage/processing capacity of the cloud center is sufficient to accommodate the generated load (unbounded).
In 2020, renewable energy accounted for 50.9\% of Germany's electricity generation\footnote{Renewables in Germany, available at \url{https://en.wikipedia.org/wiki/Renewable_energy_in_GermanyStatistics} (accessed April 2024).}. Additionally, Munich, known for its focus on solar energy, aims to transition to a fully renewable energy system by 2025\footnote{Sustainable business in Germany, available at \url{https://www.reuters.com/business/sustainable-business/germany-aims-get-100-energy-renewable-sources-by-2035-2022-02-28/} (accessed April 2024).}. For our analysis, we assume that all the networking and computing resources, except for the cloud, are supplied by the same power utility, thereby sharing a consistent renewable energy profile. Specifically, we model the energy supplied from renewable sources for each access point and its co-located fog server using a Beta distribution with parameters $\alpha=0.6$ and $\beta=0.4$.
\subsubsection{Costs}

An important consideration is the time delay involved in deploying a service. If deploying services introduce extra delay, it could significantly impact the experienced service quality. Nevertheless, deploying containers typically take less than 50 ms~\cite{kaur2017container}, while the execution duration of the proposed framework for deploying services in practical scenarios could range from several seconds to minutes. In our simulations, we set the startup delay of service containers to 50 ms.
Service costs are calculated by considering Amazon Web Services (AWS) Credits, which are determined as a percentage of monthly bill for Amazon EC2 instances. The instance chosen for analysis is t2.nano\footnote{AWS EC2, available at \url{https://aws.amazon.com/ec2/pricing/on-demand/} (accessed April 2024).}, featuring 1 vCPU, 0.5 GB of memory, and an On-Demand hourly credit of 0.0058 USD, aligning closely with the characteristics of the augmented reality service. 

\subsection{Results and Discussion}
\noindent This section presents the results obtained from four distinct sets of experiments. The initial experiment investigates the delicate balance between service provisioning costs and workload equilibrium. The subsequent two experiments delve into the intricate interplay between costs and load balance within the realms of demand variability and resource limitations, respectively. The fourth experiment scrutinizes the correlation between costs and incentivization strategies toward renewable energy resources. Notably, excluding the third experiment, all fog nodes in the other three experiments operate at full capacity, with service utilization restricted to a maximum of 90\%.

\subsubsection{Experiment 1}
\begin{figure} 
    \centering
    \includegraphics[clip, trim=0.6cm 2.8cm 1.4cm 0.3cm, width=\columnwidth]{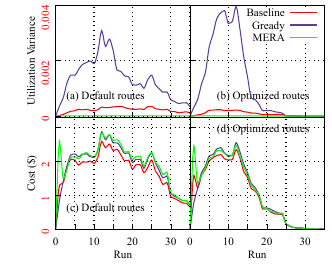}
    \caption{The influence of traffic patterns on optimization objectives reveals distinct differences in workload distribution among placement approaches, while service provisioning costs are at a similar level.}
    \label{fig:ex1gclcmethod}
    \vspace{-15pt}
\end{figure}

\begin{figure} 
    \centering
    \includegraphics[clip, trim=1.5cm 1.8cm 1.0cm 0.55cm, width=\columnwidth]{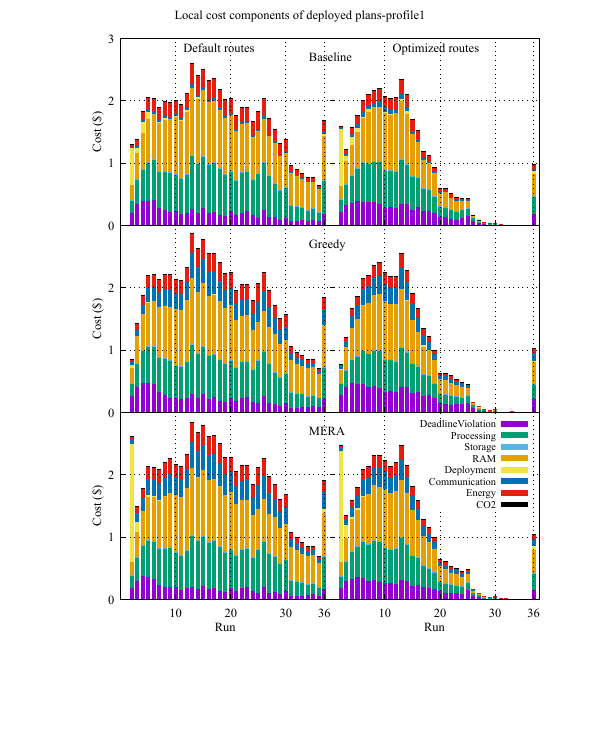}
    \caption{MERA outperforms other approaches in deadline violation, Baseline excels in communication cost, and Greedy offers the lowest deployment cost, with other costs being nearly similar. Run 36 denotes the mean value of costs across all runs for each approach.}
    \label{fig:ex1coststack}
    \vspace{-20pt}
\end{figure}

In this experiment, IoT profiles ranging from 0 to 11, as illustrated in Figure.~\ref{fig:iottraffic}, are distributed among the vehicles within the test area over a three-hour duration, comprising 36 5-min time slots. 
The association between optimization objectives (utilization variance and service provisioning cost, Figures~\ref{fig:ex1gclcmethod} and~\ref{fig:ex1coststack}) and the traffic pattern in the test area (Figure.~\ref{fig:vehtraffic}) is evident across all approaches, indicating a clear influence of traffic load on the IoT workload and cost dynamics. Nonetheless, notable differences emerge in the workload distribution offered by different placement approaches.
\par
\begin{figure} 
    \centering
    \includegraphics[clip, trim=0.3cm 2.4cm 0.3cm 0.4cm, width=\columnwidth]{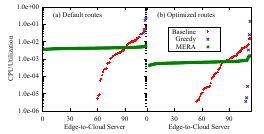}
    \caption{Greedy concentrates load on specific nodes, while Baseline distributes workload across all receivers. MERA achieves the most balanced workload distribution. Servers are sorted by utilization.}
    \label{fig:ex1util}
    \vspace{-15pt}
\end{figure}

 Figure~\ref{fig:ex1gclcmethod}a and~\ref{fig:ex1gclcmethod}b display the overall variance in CPU and memory utilization, while Figure~\ref{fig:ex1util} illustrates the CPU utilization within the edge-to-cloud nodes. These figures collectively depict the load distribution across the network for the three tested approaches.
Greedy demonstrates the highest variance by concentrating the load on specific nodes. In contrast, the Baseline, which utilizes all nodes receiving requests, achieves a more well-distributed workload with lower variance. MERA, by spreading the workload evenly across a broader range of network nodes, offers the most effective balance. 
It is evident in the scenario of the default routes, both Baseline and Greedy exhibit considerable disparities compared to MERA. The Baseline approach, on average, shows a 40-fold increase in variance, while the Greedy approach displays an even much higher imbalance. In the scenario of optimized routes, while the increase of Baseline remains notable (28 times higher), the Greedy approach demonstrates even larger variance compared to MERA. 
This observation underscores the effective workload distribution of MERA, which overcome the significant distribution inefficiencies of the Baseline and Greedy approaches. 
\par Figure~\ref{fig:ex1gclcmethod}c and~\ref{fig:ex1gclcmethod}d depict the cost of service provisioning across the three approaches, while Figure~\ref{fig:ex1coststack} dissects this cost into its constituent elements. In default scenarios, Baseline presents a modestly lower cost (10\%) compared to MERA, while Greedy shows a slight advantage (3\%) over MERA. However, in optimized route scenarios, these differences diminish, with Baseline demonstrating 6\% and Greedy 1\% lower cost compared to MERA. In the optimized scenario, the costs decrease by 45\%, 42\%, and 44\% for MERA, Baseline, and Greedy, respectively. This emphasizes the significance of uncongested traffic networks in reducing costs within ICT networks and enabling more efficient resource utilization.
\par Across both optimized and default route scenarios, the differences in storage, memory, energy, and CO2 costs between Baseline and Greedy approaches compared to MERA are lower than 2.4\%, with negligible variations observed.
Initially, MERA incurs the highest deployment cost due to distributing services across the network, while the Baseline approach experiences a smaller yet significant cost by locally deploying services on all receivers, resulting in a decrease ranging from 37\% to 48\% compared to MERA. Conversely, the Greedy approach, which deploys services on a few specific nodes, exhibits a comparatively lower cost, with a decrease of 97\% compared to MERA. The cost of deadline violation is lower in MERA, as it balances service distribution across the network, ensuring adequate processing power for each service. In contrast, the utilization of only a few nodes in Greedy leads to higher queuing and execution delays for services. Additionally, the distribution of services closer to the sources in MERA mitigates communication delays. The Baseline approach incurs a deadline cost approximately 14\% to 17\% higher than MERA, while the Greedy approach exhibits a more substantial increase, ranging from 31\% to 34\%. This underscores the effectiveness of MERA, particularly in time-sensitive applications.
The offloading of services across the network results in higher communication cost for both MERA and Greedy compared to Baseline. The MERA and Greedy approaches demonstrate a similar increase of 3\% compared to Baseline in the optimized scenario, while the difference ranges from 8\% to 9\% in the default scenario. Table~\ref{tab:costComp} in Appendix~\ref{sec:appA} offers a more detailed comparison of the costs associated with the Baseline and Greedy approaches compared to MERA.
\par The Greedy approach highlights the tendency of agents to prioritize their individual objective without coordinating towards the system-wide goal, leading to striking fragmentation over time as certain nodes handle a significant workload, while others remain underutilized. Consequently, this fragmentation leads to overloaded nodes struggling to maintain high-quality service standards, resulting in increased deadline violations and degraded overall system performance. Although the Baseline provides the lowest placement cost, it results in a better load distribution yet imbalanced. Thus, the Greedy and Baseline approaches' lack of coordination can exacerbate disparities in workload distribution, adversely impacting the system's efficiency and reliability. In contrast, MERA achieves a balanced workload at a slightly higher provisioning cost compared to the Baseline approach. However, this additional cost can be effectively mitigated by caching or storing frequently used services locally by service providers.
\subsubsection{Experiment 2}
\begin{figure} 
    \centering
    \begin{subfigure}{\columnwidth} 
    \centering
        \includegraphics[clip, trim=0.6cm 6.7cm 0.9cm 0cm, width =9.4cm]{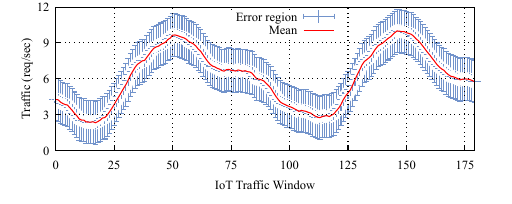}
    \end{subfigure}%
    \hspace{\fill} 
    \begin{subfigure}{\columnwidth} 
        \centering
        \includegraphics[clip, trim=0.6cm 0.6cm 0.3cm 3.4cm, width=10cm, height=6.5cm]{figures/RES/GCLC-Window-notsorted.pdf}
    \end{subfigure}%
    \caption{Traffic variation across 12-profile IoT traffic windows influences load balance and service provisioning cost. Approaches diverge notably during traffic peaks, with Greedy exhibiting the largest fluctuations}
    \label{fig:ex2window}
    \vspace{-10pt}
\end{figure}

In this experiment, a 3-hour window is initially defined, encompassing 12 consecutive 5-minute IoT profiles. Subsequently, the experiments are executed for all windows, totaling 179 windows. 
Figure.~\ref{fig:ex2window} presents the average of IoT traffic across the windows alongside the corresponding results, highlighting the influence of traffic fluctuations on the service placement goals. 
As IoT load decreases, all placement approaches show lower and more closely aligned execution costs. Conversely, with increasing load, costs diverge, especially noticeable in default route scenarios imposing heavier load on the ICT network. At the peak IoT load (window 154), the Greedy approach incurs 34\% higher costs compared to the Baseline, which decreases to 15\% in optimized routes. Comparing the cost at the peak load between MERA and the Baseline shows a 10\% increase in MERA under optimized conditions and a 15\% increase in the default scenario. Consequently, MERA demonstrates greater scalability and robustness to IoT traffic peaks compared to Greedy in terms of service provisioning cost.
\par On average, transitioning from default to optimized routes results in utilization variance reductions of 46\%, 60\%, and 51\% for MERA, Baseline, and Greedy, respectively.
This highlights a significant decrease in variability achieved with optimized traffic configurations. Regarding service provisioning cost, in the default routes scenario, MERA, Baseline, and Greedy approaches exhibit average placement costs that are 44\%, 41\%, and 46\% higher, respectively, compared to their counterparts using optimized routes. This underscores the notable reduction in costs associated with optimized traffic configurations. This discrepancy arises from the more even distribution of vehicles across the test area and quicker departure of traffic from the city via optimized routes, reducing the load on the ICT network over a shorter duration compared to default routes.

\begin{figure*} 
    \centering
        \includegraphics[clip, trim=1.2cm 9.2cm 2.5cm 0.6cm]{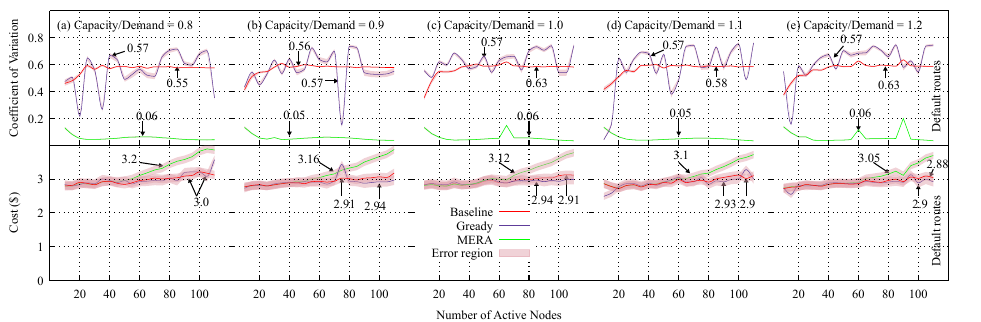}
        \caption{The influence of edge-to-cloud capacity/size on servers' utilization dispersion and service provisioning cost: Increasing capacity reduces costs, while a rise in the number of active nodes decreases utilization variation in MERA but increases it in other methods. Labels depict the mean coefficients of variation and costs for each approach.}
    \label{fig:ex3cap}
\end{figure*}

\subsubsection{Experiment 3}
This experiment assesses how network capacity and resource limitations influence service distribution and service provisioning costs. The proposed approach systematically adjusts the available capacity, ranging from 80\% to 120\% of the current demand in each run. This capacity is then allocated across a specific number of nodes (randomly selected) within the fog infrastructure, ranging from 10 to 110 nodes in increments of 5. Each experiment is repeated 10 times to ensure robustness, with each data point on the Figure~\ref{fig:ex3cap} and \ref{fig:ex3cap1} representing the average outcome derived from these 10 samples of network nodes for default and optimized routes scenarios, respectively. 
\par In the default scenario, the Greedy approach exhibits fluctuations and a rising trend as the number of active nodes increases across various capacity/demand ratios. This suggests that, despite similar capacity relative to demand, the Greedy approach faces challenges in resource allocation efficiency. Moreover, this variability intensifies with the increase in the number of nodes, as the Greedy approach opts to deploy services on a limited number of nodes among the available options. The Baseline initially shows an upward trend before stabilizing, indicating a more adaptive resource distribution compared to the Greedy approach. This leads to a more consistent server utilization pattern over time for the Baseline approach. Although fluctuations emerge with increased network capacity, possibly due to the inherent complexity of load balancing in edge-to-cloud networks, the MERA approach demonstrates a smooth decreasing trend, indicating the presence of robust load management strategies that effectively adapt to evolving network conditions.
\par The increase in fog network capacity reduces service placement costs for all approaches, benefiting both service providers and users. While the Greedy and Baseline approaches exhibit a consistent close trend across variations in the number of nodes and capacity-to-demand ratio, MERA experiences a gradual growth when the number of active nodes reaches the range of 40 to 60. In the worst-case scenario (capacity-to-demand ratio of 0.8), the cost increase can reach up to 25\% compared to Greedy and Baseline. However, this increase diminishes as the capacity-to-demand ratio increases and even falls below that of Greedy when the nodes are operating at full capacity, see Figures.~\ref{fig:ex1gclcmethod} (c,d),~\ref{fig:ex2window} (a,b), and~\ref{fig:ex4gclcmethod} (c,d). This suggests that distributing the workload across more nodes may lead to adverse effects on costs, particularly due to communication and deployment expenses. 

\subsubsection{Experiment 4}
\begin{figure} 
    \centering
    \begin{subfigure}{\columnwidth}
        \centering
        \includegraphics[clip, trim=0.3cm 3.7cm 0cm 0.7cm, width=8.2cm, height=4cm]{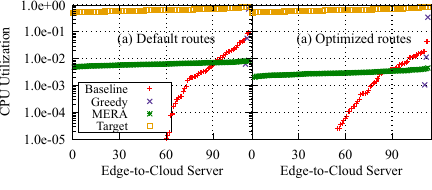}
       \end{subfigure}
    \hspace{\fill}
    \begin{subfigure}{\columnwidth} 
        \centering
        \includegraphics[clip, trim=1.1cm 4.8cm 1.5cm 0.35cm, width=8.2cm, height=4cm]{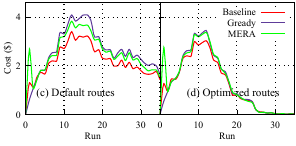}
        \end{subfigure}%
    \caption{MERA loads nodes in accordance with their renewable energy penetration rate, while Baseline deploys services locally, sacrificing load alignment for cost reduction. Greedy, on the other hand, fails to achieve superiority in either objective.}
    \label{fig:ex4gclcmethod}
    \vspace{-10pt}
\end{figure}

This experiment demonstrates how MERA prioritizes renewable energy sources over non-renewable powered servers, as illustrated by the logarithmic trends depicted in Figure~\ref{fig:ex4gclcmethod}a and~\ref{fig:ex4gclcmethod}b. MERA optimizes for this by minimizing the RMSE between the utilization and renewable power ratio of each server across the entire network.
In this figure, the orange points represent the renewable power ratio of each server, while the other points depict the utilization of those servers across different approaches. The servers are sorted by utilization. 

The upward trend observed in the MERA approach indicates that as the renewable power ratio increases, so does the utilization of the servers. This trend highlights the effectiveness of prioritizing renewable energy sources, leading to higher utilization rates. Conversely, the Baseline and Greedy approaches prioritize local deployment and minimize service provisioning costs, respectively, which lead to deviations from the target signal in these approaches.
\par Given the upward trend in incoming IoT traffic depicted in Figure~\ref{fig:iottraffic} for this experiment, Figure~\ref{fig:ex4gclcmethod}c and~\ref{fig:ex4gclcmethod}d show a notably higher provisioning cost compared to Experiment 1, where the IoT traffic remained relatively stable. Specifically, in the default scenario, we observe an increase of 29\%, 27\%, and 39\% for MERA, Baseline, and Greedy, respectively. In contrast, in the optimized scenario, the increases are relatively lower, at 22\%, 19\%, and 19\% for MERA, Baseline, and Greedy, respectively. 

\section{Conclusion}\label{sec:conc}
\noindent This paper aims to pave the way for a more efficient and sustainable smart mobility and computing ecosystem. Given the transformative potential of ITS in fostering safer and more eco-friendly transportation systems, the increasing proliferation of IoT devices within vehicles, coupled with their significant energy demands and processing costs in the context of edge-to-cloud infrastructure, poses a formidable challenge to the readiness of ICT. To address these challenges, this paper introduces a pioneering distributed service placement framework designed to dynamically manage the load of smart mobility services within the edge-to-cloud ecosystem. By optimizing QoS, service provisioning costs, workload balance, and sustainability considerations, this framework offers a novel approach to cope with the dynamic and stochastic nature of vehicular networks. Through extensive evaluation based on real-world infrastructure settings and traffic traces from Munich, the proposed framework demonstrates the potential to enhance computing resilience and empower self-adaptive ICT infrastructures. 
\par To further advance the scope and impact of this research, several avenues for future work have been identified. The framework could be extended to encompass additional dynamic edge computing applications, such as drone-based systems~\cite{qin2023coordination}.
Additionally, conducting experiments in real-world testbeds, such as the SMOTEC platform~\cite{nezami2023smotec}, could provide valuable validation and refinement of the proposed framework. Furthermore, further study into multi-modal traffic patterns, particularly those associated with the transition to low-carbon transport modalities, holds promise for yielding valuable insights into the interactions between mobility dynamics and computing resources.
\section{Acknowledgments}
\noindent This project is funded by a UKRI Future Leaders Fellowship (MR-/W009560-/1): `\emph{Digitally Assisted Collective Governance of Smart City Commons--ARTIO}'
\bibliographystyle{unsrt}
\bibliography{paper1}

\appendix
\section*{Appendix A: Supplementary Material}\label{sec:appA}
\begin{figure*}
    \centering
        \includegraphics[clip, trim=1.2cm 3.1cm 3cm 6.8cm]{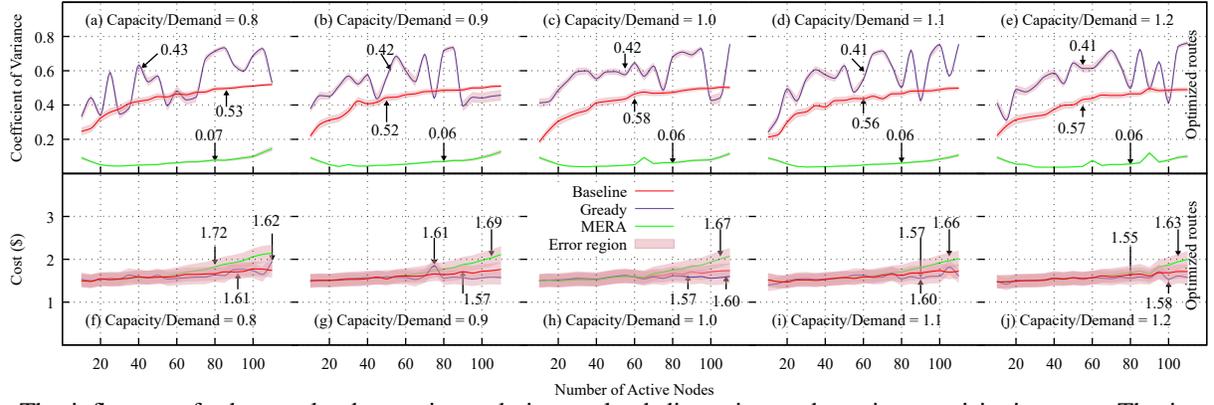}
        \caption{The influence of edge-to-cloud capacity and size on load dispersion and service provisioning cost: The increase in capacity reduces costs and converges them across all approaches, the increase in the number of active nodes increases utilization variation in all methods.}
    \label{fig:ex3cap1}
\end{figure*}

\noindent Figure.~\ref{fig:ex3cap1} illustrates the coefficient of variance and service provisioning cost in optimized routes scenario. In this scenario, all approaches exhibit reduced coefficients of variance compared to default routes, indicating that optimized routing strategies contribute to more consistent server utilization. This emphasizes the significance of efficient traffic route planning in minimizing variability and enhancing overall edge-to-cloud network performance. Particularly, the Baseline approach shows a more pronounced decrease in coefficients of variance, underscoring its effectiveness in achieving stable server utilization compared to the other approaches in the presence of optimized routes.

\par Table~\ref{tab:costComp} illustrates the changes in service provisioning cost components for the Greedy and Baseline approaches compared to MERA, highlighting both improvements (indicated by +) and deteriorations (indicated by -).
Table~\ref{tab:notation} outlines the mathematical notations utilized throughout the paper. 

\begin{table}
  \centering
  \caption{Comparison of costs and percentage differences}
  \scriptsize
  \label{tab:costComp}
  \begin{tabular}{p{1.5cm}p{1.3cm}p{1.3cm}p{1.3cm}p{1.3cm}}
    \toprule
    & \multicolumn{4}{c}{Percentage Difference with MERA (\%)} \\
    \cmidrule(lr){2-5}
    Cost& Baseline (Optimized) & Greedy (Optimized) & Baseline (Default) & Greedy (Default) \\
    \midrule
    Deadline & +14.38 & +33.79 & +16.56 & +31.71 \\
    Processing & +5.51 & -6.61 & +2.97 & -3.76 \\
    Storage & 0.00 & -0.03 & +2.13 & -2.56 \\
    RAM & -0.05 & -0.11 & -0.18 & -2.38 \\
    Deployment & -36.79 & -96.63 & -48.98 & -96.58 \\
    Communication & -75.61 & +1.81 & -90.13 & -12.16 \\
    Energy & +1.85 & +0.23 & -0.14 & -0.30 \\
    CO2 & 0.00 & 0.00 & 0.00 & 0.00 \\
    \bottomrule
  \end{tabular}
\end{table}

\tablefirsthead{%
\hline
\multicolumn{1}{l}{Notation} &
\multicolumn{1}{l}{Meaning} \\
\hline}
\tablehead{\hline
\multicolumn{1}{l}{Notation} &
\multicolumn{1}{l}{Meaning} \\
\hline}
\tabletail{%
\hline
\multicolumn{2}{r}{\small\sl continued on next column}\\
\hline}
\tablelasttail{\hline}
\topcaption{Mathematical notations}
\label{tab:notation}
\scriptsize
\begin{supertabular}{p{0.6cm}p{7.2cm}}
\hline
    C&Set of cloud nodes\\
    F&Set of fog nodes\\
    A&Set of services/vehicles\\
    M&Set of access points in the communication network\\
    $s_{i}$&Speed of vehicle i\\
    $l_{im}$&Euclidean distance between vehicle i and access point m\\
    $P_{im}$&Connection probability of vehicle i and access point m\\
    $t_{im}$&Initial connection time of i with m\\
    $t^{\mathsf{w}}_{im}$&Waiting time of i to receive reply from m\\
    $\tau$&Interval between consecutive optimization problem-solving instances (seconds)\\
    $\tau_{im}$&Connection time of vehicle i with access point m (seconds)\\
\hline
    $\delta$&Service placement plan\\
    $x_{ij}$&Binary placement decision for i on fog node j\\
    $x_{ic}$&Binary placement decision for i on cloud node c\\
\hline
    $d_{jj'}$&Propagation delay link $(j,j')$\\
    $b_{jj'}^{u}$&Average uplink transmission rate of the link $(j,j')$\\
    $b_{jj'}^{d}$&Average downlink transmission rate of the link $(j,j')$\\
    $F^{\mathsf{p}}_{j}$&Processing capacity of node j (MIPS)\\
    $F^{\mathsf{m}}_{j}$&Memory capacity of node j\\
    $F^{\mathsf{s}}_{j}$&Storage capacity of node j\\
    $u_{j}$ &Processing power utilization of node j\\
    $\mu_{j}$&Service rate of one processing unit of node j (MIPS)\\
    $n_{j}$&Number of processing units of node j\\
\hline
    $L_{i}^{\mathsf{p}}$&CPU demand of service i (million instruction per request)\\
    $L_{i}^{\mathsf{m}}$&Memory demand of service i (bytes)\\
    $L_{i}^{\mathsf{s}}$&Storage demand of service i (bytes)\\
    $\eta_i$&Desired QoS level for service i\\
    $h_{i}$&Delay threshold for service i\\
    $V_{ij}^{m}$&Delay violation percentage for service i on node j connected to m\\
    $V_{i}$&Delay violation percentage for service i on node j\\
$e_{ij}$&Delay of service i running on node j\\
    $w_{ij}$&Waiting delay for i served by node j\\
    $q_{i}$&Average size of requests of service i (bytes)\\
    $a_{i}$&Average size of responses of service i (bytes)\\
    $z_{ij}$&Arrival rate of service requests i to node j (request/second)\\
    $\zeta_{ij}^\mathsf{p}$&Traffic arrival rate of service i to node j (instruction/second)\\
    $f_{ij}$&Processing power ratio allocated to service i on node j\\
\hline   
    $P_{\delta}^{\mathsf{n}}$&Power consumption of networking equipment\\
    $P_{\delta}^{\mathsf{s}}$&Power consumption of server machines\\
    $P_{j}^{\mathsf{i}}$&Power consumption of node j in idle mode (watt)\\
    $P_{j}^{\mathsf{a}}$&Active power consumption of node j (watt)\\
    $P_{c}^{\mathsf{i}}$&Power consumption of cloud node c in idle mode (watt)\\
    $P_{c}^{\mathsf{a}}$&Active power consumption of cloud node c (watt)\\
    $p^{\mathsf{r}}_{j}$&Ratio of renewable power supplied to fog node j (server or edge/core router)\\
    $p^{\mathsf{r}}_{c}$&Ratio of renewable power supplied to cloud node c (server or switch)\\
    $p^{\mathsf{r}}_{m}$&Ratio of renewable power supplied to access point m\\
    $\theta_{c}$&PUE of cloud center c\\
    $p_{j}^{f}$&Power consumption of edge/core router j for data transfer (watt per byte)\\  
    $p_{m}^{f}$&Power consumption of access point m for data transfer (watt per byte)\\   
    $P_{\delta}^{E}$&Non-renewable power consumption of fog-cloud infrastructure\\
    $R_{c}$&Average carbon emission rate for electricity\\
\hline
    $C_{j}^{\mathsf{p}}$&Unit cost of processing at node j (\$ per million instructions)\\
    $C_{j}^{\mathsf{s}}$&Unit cost of storage at node j (\$ per byte per second)\\
    $C_{j}^{\mathsf{m}}$&Unit cost of RAM at node j (\$ per byte per second)\\
    $C_{jj'}^{\mathsf{c}}$&Communication cost of link $(j,j')$ (\$ per unit bandwidth per second)\\
    $C^{\mathsf{v}}_{i}$&Cost of deadline violation for service i (\$ per request per \%)\\
    $C^{\mathsf{r}}$&Unit cost of renewable power consumption supplied\\	
    $C^{\mathsf{n}}$&Unit cost of non-renewable power consumption supplied\\
    $C^{\mathsf{f}}$&Unit cost of carbon footprint\\
\hline
    $O_{\delta}^{\mathsf{P}}$&Cost of processing for plan $\delta$\\
    $O_{\delta}^{\mathsf{S}}$&Cost of storage for plan $\delta$\\
    $O^{\mathsf{M}}_{\delta}$ &Cost of RAM usage for plan $\delta$\\
    $O_{\delta}^{\mathsf{C}}$&Cost of communication for plan $\delta$\\
    $O_{\delta}^{\mathsf{D}}$&Cost of deployment for plan $\delta$\\
    $O_{\delta}^{\mathsf{V}}$&Cost of deadline violation for plan $\delta$\\
    $O_{\delta}^{\mathsf{E}}$&Cost of energy consumption for plan $\delta$\\
    $O_{\delta}^{\mathsf{F}}$&Cost of carbon footprint for plan $\delta$\\
    $L_{\delta}$&Local cost of service provisioning for plan $\delta$\\
    $G$&Global cost of service placement plans\\
\hline
\end{supertabular}
\normalsize

\end{document}